\documentclass[%
superscriptaddress,
nofootinbib,
nobibnotes,
amsmath,amssymb,
aps,
twocolumn,
prb]{revtex4-2}

\usepackage{lineno}
\usepackage{amsmath}
\usepackage{amssymb}
\usepackage{amsfonts}
\usepackage{graphicx}
\usepackage{dsfont}
\usepackage{bbold}
\usepackage{bm}
\graphicspath{{./imgs/}}
\usepackage{xcolor}
\usepackage{ifthen}
\usepackage{braket}
\usepackage{hyperref}
\usepackage[caption=false, subrefformat=parens]{subfig}

\usepackage[capitalize,compress]{cleveref}
\crefname{section}{Section}{Section} %
\crefname{subsection}{Section}{Section}
\crefname{equation}{Eq.}{Eqs.}
\creflabelformat{equation}{#2#1#3} %
\Crefname{equation}{Equation}{Equations}

\newcommand{\ketbra}[2]{\ket {#1} \hskip -0.8ex \bra {#2}}
\newcommand{\tr}[1][]{\ifthenelse{\equal{#1}{}}{\mathrm{Tr}\,}{\mathrm{Tr}\left[#1\right]}}

\newcommand{\graph}{\text{graph}}
\newcommand{\eff}{\text{eff}}
\newcommand{\semiemp}{\text{semi-emp.}}

\renewcommand{\Re}{\mathrm{Re}\;}
\renewcommand{\Im}{\mathrm{Im}\;}
\DeclareMathOperator{\diam}{\mathrm{diam}}
\newcommand{\diag}[2]{\vcenter{\hbox{\includegraphics[height=#1]{#2}}}}

\newcommand{\expct}[1]{\left\langle #1 \right \rangle}
\newcommand{\toy}{\text{toy}}

\begin{document}

\title{Comparing numerical methods for hydrodynamics in a one-dimensional lattice spin model}
\author{Stuart Yi-Thomas}
\email[]{snthomas@umd.edu}
\affiliation{Condensed Matter Theory Center, University of Maryland, College Park, Maryland 20742-4111, USA}
\affiliation{Joint Quantum Institute, NIST/University of Maryland, College Park, Maryland 20742, USA}
\author{Brayden Ware}
\affiliation{Joint Quantum Institute, NIST/University of Maryland, College Park, Maryland 20742, USA}
\affiliation{Joint Center for Quantum Information and Computer Science, NIST/University of Maryland, College Park, Maryland 20742, USA}
\author{Jay D. Sau}
\affiliation{Condensed Matter Theory Center, University of Maryland, College Park, Maryland 20742-4111, USA}
\affiliation{Joint Quantum Institute, NIST/University of Maryland, College Park, Maryland 20742, USA}
\author{Christopher David White}
\affiliation{Joint Center for Quantum Information and Computer Science, NIST/University of Maryland, College Park, Maryland 20742, USA}
\date{October 17, 2024}

\begin{abstract}
  In ergodic quantum spin chains, locally conserved quantities such as energy or particle number generically evolve according to hydrodynamic equations as they relax to equilibrium.
  We investigate the complexity of simulating hydrodynamics at infinite temperature with multiple methods:
  time evolving block decimation (TEBD), %
  TEBD with density matrix truncation (DMT), %
  the recursion method with a universal operator growth hypothesis (R-UOG), %
  and operator-size truncated (OST)  dynamics. %
  Density matrix truncation and the OST dynamics give consistent dynamical correlations to $t=60/J$ and diffusion coefficients agreeing within $1\%$. %
  TEBD only converges for $t \lesssim 20$, but still produces diffusion coefficients accurate within 1\%.
  The universal operator growth hypothesis fails to converge and only matches other methods on short times. %
  We see no evidence of long-time tails in either DMT or OST calculations of the current-current correlator, although we cannot rule out that they appear at longer times. We do observe power-law corrections to the energy density correlator.
  At finite wavelength, we observe a crossover from purely diffusive, overdamped decay of the energy density, to underdamped oscillatory behavior similar to that of cold atom experiments.
  We dub this behavior ``hot band second sound'' and offer a microscopically-motivated toy model.
\end{abstract}

\maketitle

\begin{figure}[t]
    \includegraphics[width=0.8\columnwidth]{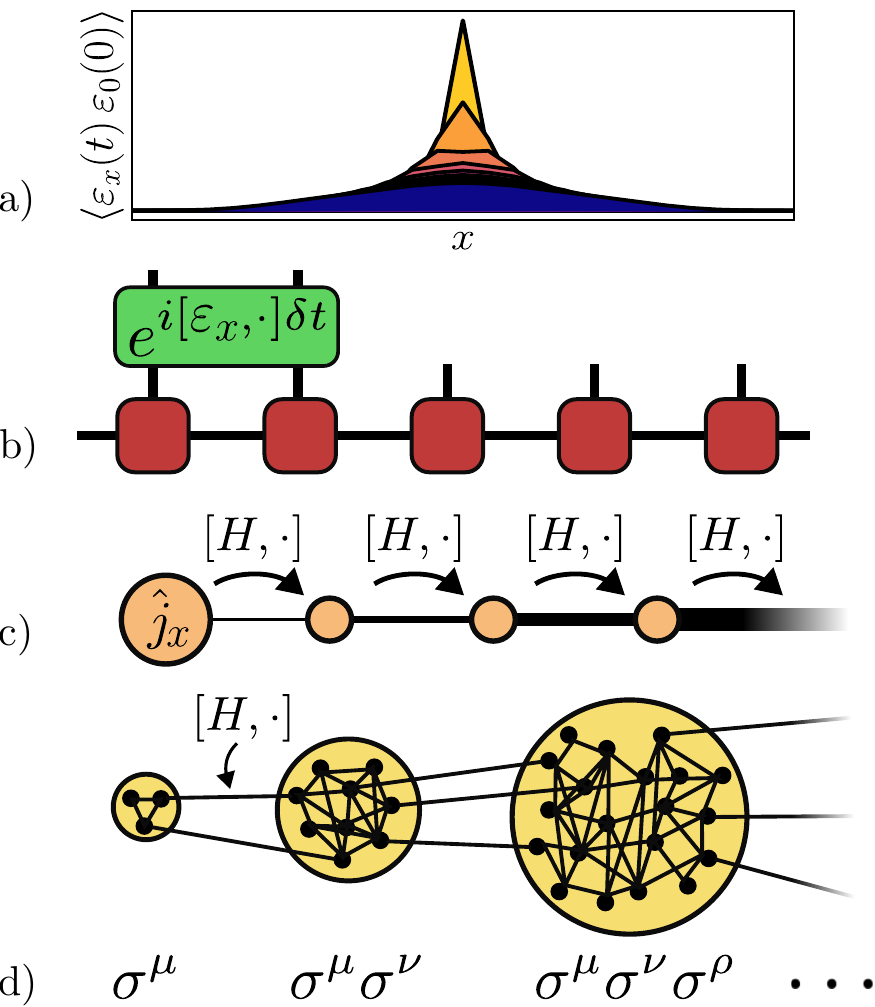}
    \caption{ \label{fig:cartoon}Illustrations of \textbf{a)} the spreading of the energy-energy correlator over time, exhibiting diffusive hydrodynamics; \textbf{b)} TEBD on a tensor network MPS; \textbf{c)} Krylov dynamics mapped onto a half-chain with increasing couplings; \textbf{d)} connectivity of the Liouvillian graph, demonstrating lower coupling between operators with different diameters. 
    }
\end{figure}

\section{Introduction}
Even at high temperature, where equilibrium states are simple, 
transport properties in strongly interacting materials can be hard to compute.
The high-temperature regime is natural in cold-atom experiments such as the optical lattice transport experiments of \onlinecite{brownBadMetallicTransport2019,wei_quantum_2022}.
These experiments engineer the system's Hamiltonian to have a low-energy subspace with dynamics governed by a Hamiltonian of interest, such as the Fermi-Hubbard model~\cite{brownBadMetallicTransport2019} or the Heisenberg model~\cite{wei_quantum_2022}.
But thermalization and transport far from the ground state are also of interest in a wide variety of systems,
from electronic materials such as cuprates and organic superconductors
\cite{zaanen2019planckian,ayres2021incoherent,poniatowski2021counterexample,spivak2010colloquium,kasahara2010evolution,sachdev2011quantum,lucasMemoryMatrixTheory2015},
which all 
display a rich finite-temperature phase diagram and non-Fermi-liquid behavior at high temperatures,
to quark-gluon plasmas~\cite{stephanov_signatures_1998,kolb_anisotropic_2000,ollitrault_anisotropy_1992,star_collaboration_experimental_2010,heinz_exploring_2015}.

Numerically exact simulation of high-temperature dynamics is difficult for large systems or long times because it requires simulation of an exponentially large Hilbert space.
But in systems that locally thermalize, one expects that sufficiently non-local operators have little effect on the dynamics of local operators, suggesting that simulations can track the local components of a Heisenberg-picture operator for long times
without tracking non-local components.
Indeed, Ref.~\onlinecite{von_keyserlingk_operator_2021} argues that in chaotic systems, one can simulate large systems to long times at a cost that is polynomial in size and time by truncating operators that are long compared to a thermalization length scale.
A number of recently proposed algorithms take advantage of this intuition, and approximately simulate high-temperature dynamics by truncating nonlocal components of Heisenberg-picture operators
\cite{white2018,rakovszky2022,kvorningTimeevolutionLocalInformation2021,parker2019,whiteEffectiveDissipationRate2021} --- 
but they differ in the details of how the non-local components are truncated.
It is unclear if these algorithms agree in their predictions and, if so, which algorithm provides the most effective truncation scheme.

To resolve this question, we use four truncations of the operator dynamics---%
the recursion method with the universal operator growth hypothesis termination (which we abbreviate as R-UOG)~\cite{parker2019},
the operator-size truncated Liouvillian graph~\cite{whiteEffectiveDissipationRate2021} (which we abbreviate as OST),
time-evolving block decimation (TEBD) of matrix product operators with the usual SVD truncation
\cite{vidal_efficient_2003,vidal_efficient_2004,zwolak_mixed-state_2004}, and
TEBD with density matrix truncations (DMT)~\cite{white2018}---%
to compute infinite-temperature transport coefficients
in the fruit-fly model for quantum dynamics, the non-integrable one-dimensional mixed-field Ising model.
Infinite-temperature transport of the mixed-field Ising model was also studied in~\onlinecite{leviatanQuantumThermalizationDynamics2017,rakovszky2022,kvorningTimeevolutionLocalInformation2021} without comparing across methods.
Because the methods differ in their underlying assumptions,
our multimethod study offers an additional level of assurance in the common results beyond what can be provided by any one method, as well as a test of those underlying assumptions.

\begin{figure}[t]
\includegraphics[width=\columnwidth]{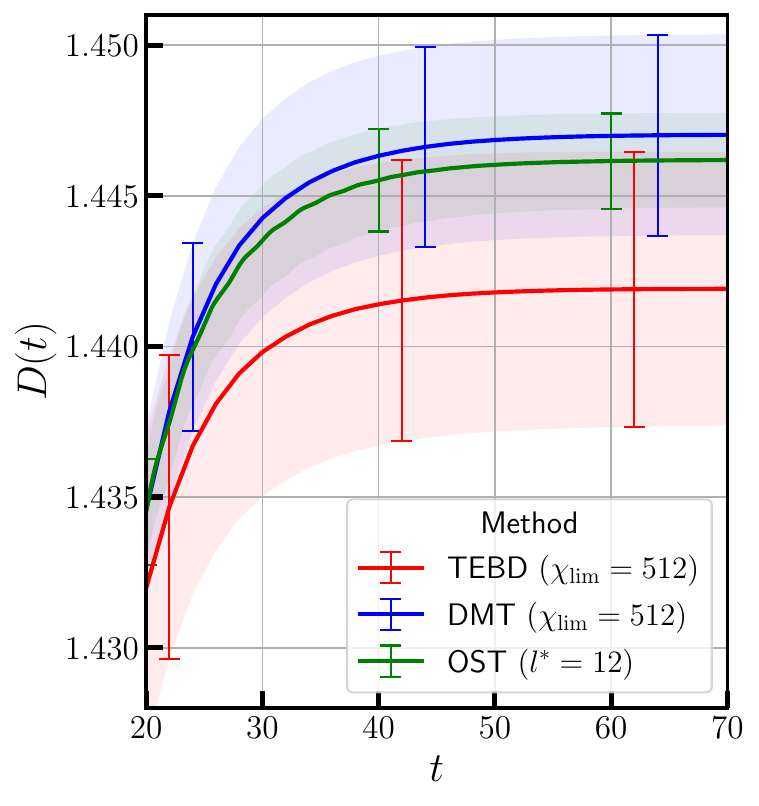}
\caption{\label{fig:Dresults} Time-dependent diffusion coefficient $D(t)$ (Eq.~\ref{eq:D-JJ}) for TEBD, DMT, and OST dynamics.
Error bars show bond dimension convergence error:
for TEBD and DMT, the absolute value of the difference between the largest bond dimensions ($|D_{\chi_\mathrm{lim} = 512}(t) - D_{\chi_\mathrm{lim} = 256}|$);
and for OST dynamics the absolute value of the difference between preservation diameters ($|D_{l_* = 12}(t) - D_{l_* = 11}(t)|$) at loss rate $\gamma_{\semiemp}$ (\cref{eq:gamma-semiemp}).
TEBD and DMT are subject to Trotter error, which turns out to be negligible (see \cref{s:trotter-convergence}).
For comparison, DAOE \cite[Fig.~3.]{von_keyserlingk_operator_2021} gives a long-time diffusion coefficient $\approx 1.4 \pm 0.01$. }

\end{figure}

We find (Figs.~\ref{fig:Dresults} and \ref{fig:dmt-tail-check}) that DMT gives current-current correlators converged in bond dimension to time $t=60$
and, consequently a diffusion coefficient that is converged to within 0.23\% at bond dimension 256.
OST agrees with these results to similar precision.
This confirms the convergence testing of \onlinecite{whiteEffectiveDissipationRate2021} and suggests that the operator chaos assumption of that work accurately models the system's behavior. %
The R-UOG, by contrast, gives current-current dynamical correlation that agree with other methods only at short times $t \lesssim 4$, when the dynamics is controlled by the Lanczos coefficients exactly calculated. Beyond that time, the recursion method fails to converge in the number of exact coefficients (Fig.~\ref{fig:uog-current}).
This sensitivity suggests that the leading, universal behavior of the Lanczos coefficients misses some essential part of the dynamics.

We additionally investigate two questions about the physics of the nonintegrable mixed-field Ising model:
the existence of hydrodynamic long-time tails and the origin of short-wavelength oscillatory modes~\cite{brownBadMetallicTransport2019,bulchandaniHotBandSound2022}.
A hydrodynamic long-time tail is a power-law (as opposed to exponential) relaxation of the current-current correlator in time,
originating from universal nonlinear corrections to the diffusion equation.
We observe that in the non-integrable mixed-field Ising model, the current-current correlator displays exponential decay across three decades in magnitude.
(We cannot rule out that the long-time tails appear at longer times.)
We do, however, observe power-law corrections to the diffusive form of the energy density correlator.
These corrections could result from nonlinear corrections to the diffusion equation, but they could also result from higher derivative terms.

Short-wavelength oscillatory modes were observed in cold atom experiments on the Fermi-Hubbard model~\cite{brownBadMetallicTransport2019}.
Ref.~\onlinecite{bulchandaniHotBandSound2022} dubbed this phenomenon ``hot band sound'' and observed it in numerical simulations of a kinetically constrained model with particle number conservation.
We give numerical evidence that the fruit-fly non-integrable Hamiltonian, which does not conserve particle number, also has short-wavelength oscillatory modes in the energy density.
We dub these modes ``hot band second sound'' and we give a physical picture for how these oscillatory modes emerge.

The paper is organized as follows.
In Sec.~\ref{s:model-phenom}, we discuss our model (the Ising model model with transverse and longitudinal fields)
and the phenomenology we expect:
diffusion at leading order in frequency and wavelength, with higher-order corrections.
In Sec.~\ref{s:methods} we describe the methods we use.
In Sec.~\ref{s:JJ} we give the results of the four methods for the current-current correlator.
And in Sec.~\ref{s:osc} we describe the model's short-wavelength ``hot band second sound'' phenomenon.

\section{Model and phenomenology}\label{s:model-phenom}

\subsection{Model}

We consider spin chains described by nonintegrable, translationally invariant, parity-symmetric, and time-independent Hamiltonians with no additional symmetries. Such systems have just one local conserved quantity, the energy density.
We use the spin-1/2 Ising model with both transverse and longitudinal fields
\begin{equation}
    H = \sum_{i}  4 J S^{z}_{i}S^{z}_{i+1} + 2 g_{z}S^{z}_{i} + 2 g_{x}S^{x}_{i}.\label{eq:ising-model}
\end{equation}
where $S^a = \frac 1 2 \sigma^a$ are spin operators.
This model is well-studied in the context of hydrodynamics%
~\cite{kim_ballistic_2013,white2018,rakovszky2022,whiteEffectiveDissipationRate2021}
and follows the ETH in a strong sense~\cite{kim_testing_2014}.
In this paper, we set $J=1$, $g_{x}=1.4$, and $g_{z}=0.9045$ for consistency with other work on numerical methods for hydrodynamics~\cite{white2018,rakovszky2022}. These particular values of the parameters are chosen to ensure that the model is far from the integrable lines $g_z=0$ or $g_x=0$, and similarly far from the near-integrable physics that occurs when $g_x$ or $g_z$ is large. Our results should not change qualitatively as long as $J$, $g_x$, and $g_z$ are all of similar magnitude.

In matrix product operator (MPO) simulations, we work with large but finite systems ($L = 256$), stopping the simulations before the bond dimension near the boundary becomes greater than 1.
R-UOG and the OST dynamics can be formulated directly in Fourier space, so in those simulations we work with formally infinite systems. 

\subsection{Phenomenology}\label{ss:phenomenology}

Ergodic quantum systems, such as the non-integrable spin chains under study in this paper, approach thermal equilibrium under their own dynamics.
When the dynamics conserves some charge, the approach to equilibrium is generically described at the largest scales by \textit{hydrodynamics}.
Hydrodynamics is a classical effective description derived by assuming that the state is locally similar to the maximum entropy state with constrained values of the density of conserved charges.
This assumption only makes sense as a coarse-grained, approximate description~\cite{landau2013fluid, doyonLectureNotesGeneralised2020}; the hydrodynamic equations are irreversible, unlike the true microscopic dynamics. 
(One goal of~\onlinecite{von_keyserlingk_operator_2021}  and~\onlinecite{whiteEffectiveDissipationRate2021} is to provide a microscopic justification for this irreversibility.)

Consider a spin chain with a single local conserved quantity.
In this paper, the conserved quantity will be an energy density operator
\begin{equation}
\label{eq:energy-density}
        \hat \varepsilon_x = 4 J S^{z}_{x}S^{z}_{x+1} + g_{z}(S^{z}_{x} + S^z_{x+1}) + g_{x}(S^{x}_{x} + S^x_{x+1}).
\end{equation}
The corresponding local current operator $\hat{j}_x$ is defined via the local conservation law 
\begin{align}\label{eq:cl-continuity}
  \partial_t \hat{\varepsilon}_x = -\Delta \hat{j} = \hat{j}_{x-1} - \hat{j}_x \;.
\end{align}
which for energy density in \cref{eq:energy-density} yields\footnote{There is a gauge ambiguity in the definition of $\hat \varepsilon_x$ and $\hat j_x$, as rewriting $\hat \varepsilon_x \to \hat \varepsilon_x + \hat a_{x+1}-\hat a_x$ for any local operator $\hat a_x$ yields the same total energy $\hat H$ up to a boundary term. Following Ref.~\onlinecite{nardisHydrodynamicGaugeFixing2023}, we assume that $\hat \varepsilon_x$ is chosen in a translationally invariant and PT-symmetric manner.}
\begin{equation}
\hat j_x = 4g_x \left(S^y_x S^z_{x+1} - S^z_{x-1} S^y_x \right)
\end{equation}
The assumption underlying hydrodynamics is that the system's state, including currents,
is entirely determined by the values of the local conserved charge density; thus
$j(x)=\langle \hat j_x \rangle$ is a functional of the profile of $\varepsilon(x) = \langle \hat \varepsilon_x \rangle$~\cite{landau2013fluid, doyonLectureNotesGeneralised2020}.
Because the dynamics are local, we can expand $j$ in gradients of $\varepsilon$:\footnote{Even-order derivatives disappear because the current is odd under inversion symmetry.}
\begin{align}\label{eq:j-grad-expansion}
  \begin{split}
    j =-D \partial_x \varepsilon + &F_{1}\ \partial_x^3\varepsilon + F_{2}\ \partial_x^5 \varepsilon\dots \\
        +&F_{3}\ \varepsilon \partial_x \varepsilon  + F_{4}\ \varepsilon \partial_x^3 \varepsilon + \dots \\
        +&F_{5}\ \varepsilon^2 \partial_x \varepsilon + F_6 \varepsilon^2\partial_x^3 \varepsilon + \dots \\
        +&F_{7}(\partial_x \varepsilon)^3 + \dots \\
  \end{split}
\end{align}
At leading order in momentum and in the variation in $\varepsilon$, the gradient expansion becomes $j =-D \partial_x \varepsilon$, so the energy $\varepsilon$ follows the diffusion equation
\begin{align}\label{eq:diffusion}
  \partial_t \varepsilon = -D \partial_x^2 \varepsilon.
\end{align}

Transport of the energy density $\varepsilon$ can be studied with the two-point correlation function
\begin{equation}
    C^{\varepsilon\varepsilon}(x, t) = \left \langle \hat{\varepsilon}_x(t) \hat{\varepsilon}_0(0) \right \rangle \equiv \text{Tr}\left[\hat{\varepsilon}_x(t) \hat{\varepsilon}_0(0) \rho_{\text{eq}}\right], \\
\end{equation}
where the expectation value is taken with respect to an equilibrium state $\rho_{\text{eq}}$. In linear response, this correlation function measures the spread of an infinitesimally small bump in energy density initially located at $x=0$.
We can define a variance
\begin{align}\label{eq:variance}
    V(t) &= \frac{ \sum_x x^2 C^{\varepsilon\varepsilon}(x, t)}{\sum_x C^{\varepsilon\varepsilon}(x, t)} .
\end{align}
The denominator
\begin{equation}
\label{eq:nu}
     \nu \equiv \sum_x C^{\varepsilon\varepsilon}(x, t) = \frac 1 L \mathrm{Tr} \, H^\dag H
\end{equation}
is time-independent by the conservation of energy. For the Hamiltonian in \eqref{eq:ising-model}, $\nu=J^2+g_z^2+g_x^2$.

In a system where the energy $\varepsilon$ undergoes diffusion, $C^{\varepsilon\varepsilon}(x, t)$ asymptotically takes the form of a Gaussian with a spatial variance $V(t)$ that increases linearly in time:
\begin{subequations}
\begin{align}%
  C^{\varepsilon\varepsilon}(x, t) &\sim \frac 1 {\sqrt{2\pi V(t)} } \exp \left[-\frac {x^2}{2V(t)} \right],  \\
    V(t) &\approx 2 D t\;. \label{eq:var-t}
\end{align}
\end{subequations}
Even for diffusive systems, Eq.~\ref{eq:var-t} only holds in the long-time limit.
To understand how the system approaches that long-time limit
(and to treat systems that may be super- or sub-diffusive),
we define a time-dependent diffusion coefficient 
\begin{equation} \label{eq:Dt}
D(t) = \frac12 \frac{d}{dt} V(t)
\end{equation}
using the variance in Eq.~\ref{eq:variance}.
In the special case of a diffusive systems, $D(t)$ approaches a constant $D$ in the long-time limit.
In general---that is, without assuming diffusion---the conservation law \eqref{eq:cl-continuity} relates the time-dependent diffusion coefficient $D(t)$ to the autocorrelation of the total current operator $\hat{J} = \sum_x \hat{j}_x$~\cite{steinigewegDensityDynamicsCurrent2009}, assuming time- and spatial-translation invariance:
\begin{subequations} \label{eq:D-JJ}
\begin{align}
D(t) &= \frac{1}{\nu} \int_0^t dt' \sum_x \left \langle \hat{j}_x(t) \hat{j}_0(0) \right \rangle 
\label{eq:current-grns}
\\
 &=  \frac{1}{L \nu} \int_0^t dt' \left \langle \hat{J}(t') \hat{J}(0) \right \rangle.
\label{eq:current-grns2}
\end{align}
\end{subequations}
Much of this paper treats the current autocorrelation function to calculate $D(t)$. 
Derivations of these correlators, as well as a comparison of the resultant values of $D(t)$, are found in Appendix \ref{sec:correlator-comparison}.
Neither the value of $D$ nor the other coefficients in \cref{eq:j-grad-expansion} are straightforwardly derived from the microscopic equations of motion. 

For finite times, $D(t)$ may be sensitive to the nonlinear-in-$\varepsilon$ corrections $F_{3}, F_4, F_5, \dots$ of the gradient expansion of the current \eqref{eq:j-grad-expansion}.
In particular, 
these nonlinear corrections can lead to \textit{long-time tails}~\cite{mukerjee_statistical_2006}:
roughly, algebraic decay of $D(t)$ to its long-time value.
When computing the diffusion coefficient via the correlator $\langle J(t) J(0)\rangle$, as in \cref{eq:D-JJ},
many of the terms in the gradient expansion given by \cref{eq:j-grad-expansion} disappear.
In particular, terms of the form $\varepsilon^n \partial_x \varepsilon^l$ become total derivatives.
Since $\lim_{x \to \pm \infty } \langle \varepsilon(x) \rangle = 0$, these must integrate to zero.
For systems with a single conserved charge,
the lowest-order term that is not of this form is $(\partial_x \varepsilon)^3$;
in \cref{eq:D-JJ} the term $(\partial_x \varepsilon)^3$ does not become an integral of a total derivative,
though we cannot rule out the possibility that it disappears for other reasons.%
\footnote{\onlinecite{mukerjee_statistical_2006} analyzed systems with multiple conserved quantities $q_\alpha$.
In that context, the first nonlinear correction that is not an integral of a total derivative is a second order term $j_\alpha \propto \dots + E_{\alpha\beta\gamma} q_\beta \partial_x q_\gamma$.}
(One of our results is that if $(\partial_x \varepsilon)^3$ contributes to the integral in (10),
its effect is small.)

These long-time tails do not exhaust the system's potential finite-wavelength and finite-frequency dynamics.
To probe short-wavelength physics beyond leading-order hydrodynamics, 
we also analyze the Fourier transform of the charge-charge correlator
\begin{align}\label{eq:energy-grns}
  C^{\varepsilon\varepsilon}(k,t) =  \expct{\varepsilon(k,t) \varepsilon(k,0)} = \sum_x e^{-ikx} C^{\varepsilon\varepsilon}_x(t)\;.
\end{align}
For a strictly diffusive system $C^{\varepsilon\varepsilon}(k,t) \propto e^{-Dk^2t}$,
but non-diffusive short-wavelength and high-frequency dynamics can lead to other forms.
This correlator measures the system's response to a drive, so many experiments can probe it or its time Fourier transform~\cite{forster_hydrodynamic_1975}.
It is also used to analyze the quantum gas microscope experiment of \onlinecite{brownBadMetallicTransport2019,Hild2014}.

\section{Methods}\label{s:methods}

To compute the correlators discussed in the previous sections, we use several methods for evolving operators. 
Two of the methods---TEBD and TEBD with density matrix truncation (DMT)---%
approximate the evolving operator as a matrix product operator (MPO) in a real space basis, as illustrated in \cref{fig:cartoon}(b).
DMT differs from TEBD in that it guarantees that local expectation values are preserved during truncation.
These methods are further discussed in Sec.~\ref{sss:tebd} and
Sec.~\ref{sss:dmt} respectively.

The other two methods, the recursion method with universal operator growth hypothesis (R-UOG, Sec.~\ref{ss:uog}) and the operator size truncated dynamics (OST, Sec.~\ref{ss:liouv}), 
both represent Heisenberg dynamics as single-particle hopping on a graph.
The recursion method maps the Heisenberg dynamics of the operator to single-particle hopping on a half-infinite chain with hopping coefficients---called Lanczos coefficients---computed via repeated commutators with the Hamiltonian, as illustrated in \cref{fig:cartoon}(c).
The truncated Liouvillian graph likewise maps the Heisenberg dynamics of the zero-momentum current operator to single-particle hoppings on a graph,
but now the graph has vertices labeled by Pauli strings and edges specified by the Hamiltonian, as illustrated in \cref{fig:cartoon}(d).
The dynamics among Pauli strings with small diameter is treated exactly, with the coupling to larger operators replaced by an effective dissipative boundary, as proposed by one of the authors in Ref.~\onlinecite{whiteEffectiveDissipationRate2021}.

\subsection{Matrix product operator methods}\label{ss:mpo-methods}

To compute the current Green's function \eqref{eq:current-grns}
using TEBD simpliciter (Sec.~\ref{sss:tebd}) or TEBD with density matrix truncation (DMT, Sec.~\ref{sss:dmt}),
we construct an matrix product operator (MPO) representation of $\hat j(t)=e^{-iHt}\hat j(0) e^{iHt}$.
\Cref{eq:current-grns}
then gives the Green's function for time $t$. For long times, we use time-doubling (\cref{s:time-doubling}) to calculate results up to $2t$ using the same MPO.
In Sec.~\ref{s:osc} we compute the energy density Green's function \eqref{eq:energy-grns} analogously using DMT.

\subsubsection{TEBD}\label{sss:tebd}
Time-evolving block decomposition~\cite{vidal_efficient_2003,vidal_efficient_2004} applies a Trotterized time evolution to a matrix product state or matrix product operator~\cite{zwolak_mixed-state_2004}.
(For an exhaustive review of matrix product state methods, see~\onlinecite{schollwoeck_density-matrix_2011}).
We use a fourth-order Trotterization~\cite[Eq.~32 ($m = 13$, type SL, $\nu= 3$)]{barthel2020}.

Each gate application increases the MPO bond dimension,
but the resulting state may have a good approximation with lower bond dimension.
TEBD chooses the approximate MPO that minimizes the $L_2$ distance $\tr (\rho - \rho_{\text{approx}})^2$.
TEBD truncation generically changes the system's conserved quantities and currents. 

\subsubsection{DMT}\label{sss:dmt}

\textit{Density matrix truncation} (DMT)~\cite{white2018} uses the same limited bond dimension MPO approximation as TEBD, 
while guaranteeing that local expectation values%
---including conserved quantities and currents---%
are unchanged in the truncation process.
It does this by performing a gauge transformation on the virtual space
that separates correlations on length scales less than some controllable length from longer-range correlations,
and discarding small principal components of the long-range correlations.
We describe the method in some detail in App.~\ref{app:dmt-details}.
DMT as presented in \onlinecite{white2018} was designed for Schr\"odinger dynamics of density matrices and uses the fact that those density matrices have nonzero trace.
We use a variant of DMT slightly modified for Heisenberg evolution of traceless operators (\cref{app:dmt-details}).

\subsection{Recursion method with universal operator growth hypothesis termination (R-UOG)}\label{ss:uog}

The recursion method is a general tool for dynamics that derives from the Lanczos tridiagonalization algorithm~\cite{mattisHowReducePractically1981,viswanathRecursionMethodApplication1994}. We follow the implementation of the method in Ref.~\onlinecite{parker2019}.

As in all Lanczos-derived algorithms, the computation begins by the construction of an orthonormal basis for the space of 
vectors generated by the repeated action of an operator.
We consider the Hamiltonian dynamics of operators generated by the Liouvillian superoperator $\mathcal{L} = [H, \cdot]$ and initial operator $\mathcal{O}_0 = J$, the energy current.
We use the trace inner product to define orthonormality.
The recursion method then constructs an orthonormal basis for the space of operators spanned by repeated commutations of the Hamiltonian with the current operator:
\begin{multline}
\{\mathcal{O}_0 = J,\; \mathcal{O}_1=\mathcal{L}(\mathcal{O}_0),\; \mathcal{O}_2=\mathcal{L}(\mathcal{O}_1)\ldots \} \\
=\{J,\; [H, J],\; [H, [H, J]],\; \ldots\}.
\end{multline} 
The orthonormal basis $\{B_k\}$ is generated by the Gram-Schmidt procedure, which can be expressed through the recurrence relation
\begin{align} \label{eq:recursion}
b_k &= \left(\text{Tr}\, A_k^{\dagger} A_k\right)^{1/2} \nonumber \\
B_k &= b_k^{-1} A_k \nonumber \\
A_{k+1} &= \mathcal{L} B_{k} - b_{k} B_{k-1}.
\end{align} 
The recursion is initialized with $A_0 = \mathcal{O}_0 = J, B_{-1} = 0.$
In the basis $\{B_k\}$, the Liouvillian takes a tridiagonal form with coefficients given by $b_k$:
\begin{equation}
  \mathcal{L}_{\text{eff}} = \begin{bmatrix}
0 & b_1 &  & &  \\
b_1 & 0 & b_2 & & \\
& b_2 & 0 & \ddots & \\
 & & \ddots & \ddots &  \end{bmatrix}.
\end{equation}
Once $b_k$ and $B_k$ are computed, the time evolution 
\begin{equation}
J(t) = e^{-i \mathcal{L} t} J
\end{equation}
can be computed using the effective tridiagonal Louivillian $\mathcal{L}_{\text{\eff}}$.
This maps the evolution problem to that of a particle hopping on a half-infinite,
one dimensional chain, as depicted in \cref{fig:cartoon}(c), with hopping coefficients $b_k$.

We work directly in the thermodynamic limit, where the recursion never ends.
The recursion coefficients can be computed with finite memory
using a sparse, momentum space representation of the operators that stores only non-zero coefficients in a Pauli operator basis.
The computational cost for each successive step grows exponentially in the Lanczos order $k$
due to the combinatorial explosion of the number of terms in the repeated commutators.
In practice, a maximum of $k^* \sim 20\mbox{--}30$ Lanczos coefficients can be exactly computed.

The recursion method handles this limitation by choosing a termination of the half-infinite chain appropriate to 
the dynamics at hand~\cite{viswanathRecursionMethodApplication1994}. A hard-wall termination at $k^*$ is a particularly
bad choice: due to reflections off the end of the chain, such a time-evolution predicts unphysical revivals of the current.
Ref.~\onlinecite{parker2019} proposes instead to continue the chain indefinitely, using an ansatz---the \textit{universal operator growth hypothesis}---to approximate the  Lanczos coefficients $b_k$ for $k>k^*$.
They give persuasive evidence that generic, ergodic, one-dimensional spin chains have coefficients $b_k$ that scale asymptotically as $b_k \sim k/ \log k$.

We therefore fit the computed ($k\leq k^*$) Lanczos coefficients to a form
\begin{equation}
b_k = a k /\log k + c\;,
\end{equation}
and compute the dynamics using Lanczos coefficients
\begin{equation}
  b'_k =
  \begin{cases}
    b_k & k \le k^* \\
    a k /\log k + c& k^* < k \le K
  \end{cases}
\end{equation}
with the fit parameters $a, c$ extracted from the first $k^*$ coefficients. 
For our purposes, $K$ is taken to be any value that is large enough to ensure the operator does not reach the boundary within the time of the simulation. As an alternative, one can use a Green's function termination at site $K$ to appropriately account for the dynamics on the remainder of the chain~\cite{viswanathRecursionMethodApplication1994, parker2019}.

\subsection{Operator-size truncated Liouvillian graph with a priori decay (OST)}\label{ss:liouv}

Instead of mapping Heisenberg dynamics to single-particle hopping on a one-dimensional half-infinite chain,
the picture of~\onlinecite{whiteEffectiveDissipationRate2021}
maps it to a more complicated graph,
whose vertices are Pauli strings and whose edges are given by matrix elements of the Liouvillian.
This graph is called the Liouvillian graph;  it was introduced in \onlinecite{altman_ehud_computing_2018}.
In the OST dynamics of \onlinecite{whiteEffectiveDissipationRate2021},
the structure of this graph, together with a chaos assumption on the dynamics of operators that act nontrivially on many sites,
leads to a strongly-interacting analog of Boltzmann's Stosszahlansatz~\cite{ehrenfest_conceptual_2002,gottwald_boltzmanns_2009}---his assumption that correlations between particle velocities can be ignored.
This in turn leads to a tractable non-Hermitian effective model of the dynamics of short operators.

The Heisenberg dynamics of an operator $A = \sum_{ \mu} A_{ \mu} \sigma^{ \mu}$, where $\sigma^\mu$ are Pauli strings, is
\begin{align}
  \frac d {dt} \sum_{ \mu} A_{ \mu} \sigma^{ \mu} = \sum_{ \mu}  i[H,\sigma^{ \nu}]A_{ \mu}\;.
\end{align}
This is exactly equivalent to a single particle hopping between sites labeled by the Pauli strings $\sigma^{ \mu}$ with amplitudes given by the commutator:
\begin{align}\label{eq:ham-graph}
  \begin{split}
    H_{\graph} &= \sum_{ \mu  \nu} L_{\mu\nu}c^\dagger_{ \mu} c_{ \nu}\\
    L_{\mu\nu} &= 2^{-L}\,\tr \big[ \sigma^{ \mu}[ H, \sigma^{ \nu}] \big]\;.
  \end{split}
\end{align}
For a local Hamiltonian almost all of the $L_{\mu\nu}$ are $0$,
so it is useful to think of the Hamiltonian \eqref{eq:ham-graph} as hopping on a graph
whose vertices are the Pauli strings
and whose edges are pairs $(\mu,\nu)$ with $L_{\mu\nu} \ne 0$.\cite{altman_ehud_computing_2018}.

The Liouvillian graph has a natural subgraph structure.
Any one Pauli string is connected to many other Pauli strings of the same length,
but very few larger or smaller strings.
More explicitly, define the \textit{diameter} of a Pauli string $\diam \sigma^\mu$
to be the distance between the first site on which it acts nontrivially and the last%
---that is, the diameter of the Pauli string's support.
Write
\begin{align*}
  \mathcal G_l = \{\sigma^\mu : \diam \sigma^\mu = l\}\;;
\end{align*}
where $\mathcal G_l$ is the \textit{pool of diameter $l$}.
If the Hamiltonian generating the Liouvillian graph is local,
there are $O(l)$ Hamiltonian terms that can commute with some $\sigma^\mu \in \mathcal G_l$
to produce other Pauli strings in $\mathcal G_l$,
but only $O(1)$ terms that can produce Pauli strings in $\mathcal G_{l\pm 1}$.
The pools $\mathcal G_l$ are therefore tightly intraconnected, but loosely interconnected.

Moreover,
if the physical Hamiltonian is not integrable, the dynamics of the graph Hamiltonian
restricted to a pool $\mathcal G_l$ is chaotic.
Its eigenoperators have support on all the Pauli strings in $\mathcal G_l$,
its eigenenergies display GOE or GUE level-spacing statistics,
and its density of states is (approximately) Gaussian.
For large $l$, then, the dynamics of the pool $l$ are completely characterized by the Gaussian Green's function
\begin{align}\label{eq:pool-greens}
  G_{\lambda\lambda;l}(t) &\equiv \braket{\lambda|\exp\left[-iH|_{\mathcal G_l}\,t\right]|\lambda} \nonumber \\
  &\approx e^{-\gamma_l^2t^2}
\end{align}
where $H|_{\mathcal G_l}$ is the restriction of the graph Hamiltonian \eqref{eq:ham-graph} to the pool $\mathcal G_l$,
$\lambda$ is some operator in $\mathcal G_l$,
and $\gamma_l$ is the width of the density of states
\begin{align}
  \gamma_l^2 = \frac 1 {|\mathcal G_l|} \tr H|_{\mathcal G_l}^2\;.
\end{align}
The Green's function \eqref{eq:pool-greens} is broadly independent of the operator $\lambda$ if $H|_{\mathcal G_l}$ is chaotic, as it is for the mixed-field Ising model.
For the mixed-field Ising model \eqref{eq:ising-model} this is $ \gamma_l^2 = l(J^2 + g_x^2 + g_z^2 )$.
Ref.~\onlinecite{whiteEffectiveDissipationRate2021} found that the dynamics on all pools with diameter $l$ or larger, i.e. $\bigcup_{l' \ge l} \mathcal G_{l'}$,
is better characterized by a semi-empirical decay rate
\begin{align}
  \label{eq:gamma-semiemp}
  \gamma_\semiemp = \sqrt{(l+1)(J^2 + g_x^2 + g_z^2 )}\;.
\end{align}
The physical picture behind \eqref{eq:pool-greens} is that our single particle can hop onto a site in the pool $\mathcal G_l$
and stay there for a time $\sim \gamma_l^{-1}$, during which it can hop back to the pool, say $\mathcal G_{l-1}$, from which it came.%
\footnote{
  We assume the particle is very unlikely to hop back down to $\mathcal G_{l-1}$ after it has left the first site in $\mathcal G_{l-1}$ it hopped to.
  \onlinecite{whiteEffectiveDissipationRate2021} offers numerical evidence that this is the case and intuition for how it comes about;
  \onlinecite{von_keyserlingk_operator_2021} offers more careful combinatorial arguments about a related question.
}

This physical picture suggests that we pick some $l$, call it $l_*$, and replace the dynamics on operators longer than
$l_*$ by a memory kernel~\cite{zwanzig_memory_1961,mori_transport_1965} (see \onlinecite[chapter 5]{forster_hydrodynamic_1975} for a helpful introduction) given by the Green's function \eqref{eq:pool-greens}.
This would lead to an unpleasant integro-differential equation.
Instead, we (following \onlinecite{whiteEffectiveDissipationRate2021}) replace the Gaussian Green's function \eqref{eq:pool-greens} by an exponential memory kernel.
The result is a non-Hermitian effective Hamiltonian
\begin{align}\label{eq:graph-eff}
  \begin{split}
  H_{\graph;\eff}(\gamma,l_*)  &= \sum_{%
  \bm \mu, \bm \nu \in \mathcal G_{\leq l^*} %
} L_{\bm\mu\bm\nu}c^\dagger_{\bm\mu} c_{\bm\nu} + \mathrm{h.c.}\\
    &- i\gamma \sum_{%
    \bm \nu \in \mathcal G_{l^*+1}%
    } c^\dagger_{\bm \nu} c_{\bm \nu}\;.
  \end{split}
\end{align}
In principle $\gamma$ is set by the semi-empirical prediction \eqref{eq:gamma-semiemp},
but we will find it enlightening to consider how the behavior of the effective model depends on the decay rate $\gamma$,
so we leave it as a parameter.

As long as the underlying physical Hamiltonian is translation invariant, the effective graph Hamiltonian can be directly constructed in Fourier space.
We do so throughout this paper.
We use Krylov subspace methods to simulate the dynamics generated by the effective Hamiltonian \eqref{eq:graph-eff}~\cite{juthoKrylovKitJl2023}.

The OST dynamics is motivated in part by DAOE~\cite{rakovszky2022}, but differs substantially in how it treats the operator Hilbert space.
DAOE keeps the whole operator Hilbert space, but modifies the evolution by adding an artificial dissipation on large-weight operators.
Because DAOE uses matrix product operators as the underlying data structure,
it can represent operators in a manifold (the low-bond-dimension manifold)
that includes a complete basis for that Hilbert space;
the artificial dissipation keeps the system in or near that low-bond dimension manifold.
OST dynamics, by contrast, truncates the Hilbert space---it discards operators of diameter greater than $l_*$---and models the effect of those operators on the small-diameter operators by a decay.

The OST dynamics differs from the dynamics of \onlinecite{altman_ehud_computing_2018} in how it truncates the Liouvillian graph. 
The dynamics of \onlinecite{altman_ehud_computing_2018} imposes an absorbing boundary condition %
at a fixed graph distance from an initial seed operator.
The OST dynamics \cite{whiteEffectiveDissipationRate2021} instead puts the absorbing boundary at a natural subgraph partition of the Liouvillian graph,
and motivates the absorbing boundary condition using the chaos properties of the subgraph dynamics.

\section{Current-current correlator}\label{s:JJ}

\subsection{Exponential decay of current-current correlator}

\begin{figure}[t]
  \includegraphics[width=\columnwidth]{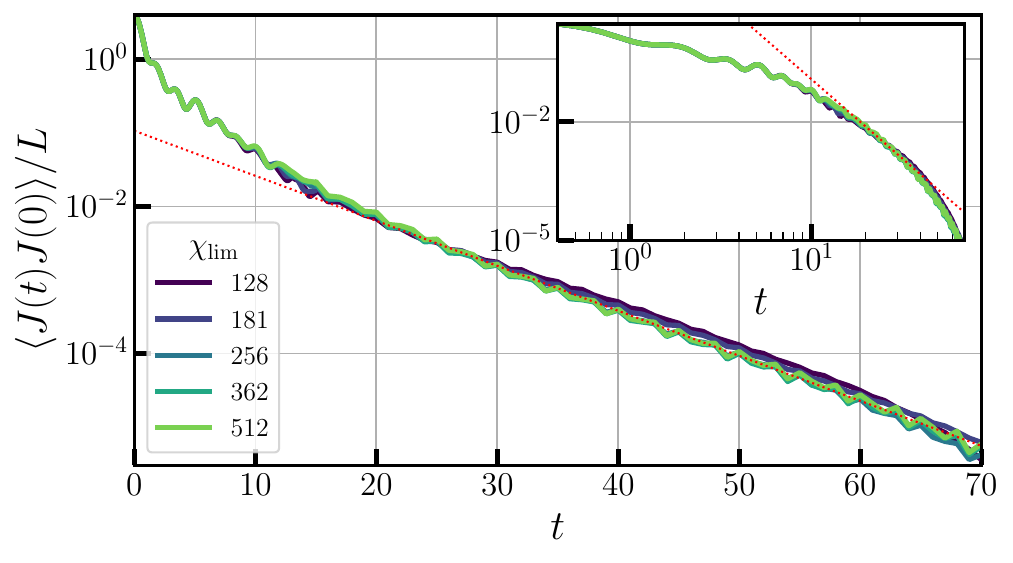}
  \caption{
  \label{fig:dmt-tail-check}
  Current-current correlator computed with DMT, across bond dimension caps $\chi_\mathrm{lim}$. The correlator displays approximately three decades of exponential decay (for $20 \lesssim t \lesssim 70$) at decay rate $0.14$ (red dotted line).
  The inset demonstrates the lack of a predicted \cite{michailidis2023,mukerjee_statistical_2006} $\sim t^{-4}$ (red dotted line)  power law via a logarithmic time axis.
    This plot shows Trotter step $\tau = 0.05$ for $t<15$ and $\tau=0.5$ with time-doubling (see \cref{s:time-doubling}) for $t>15$; Fig.~\ref{fig:trotter-convergence} shows convergence in Trotter step.
  }
\end{figure}

\begin{figure}[t!]
  \centering
  \includegraphics[width=\columnwidth]{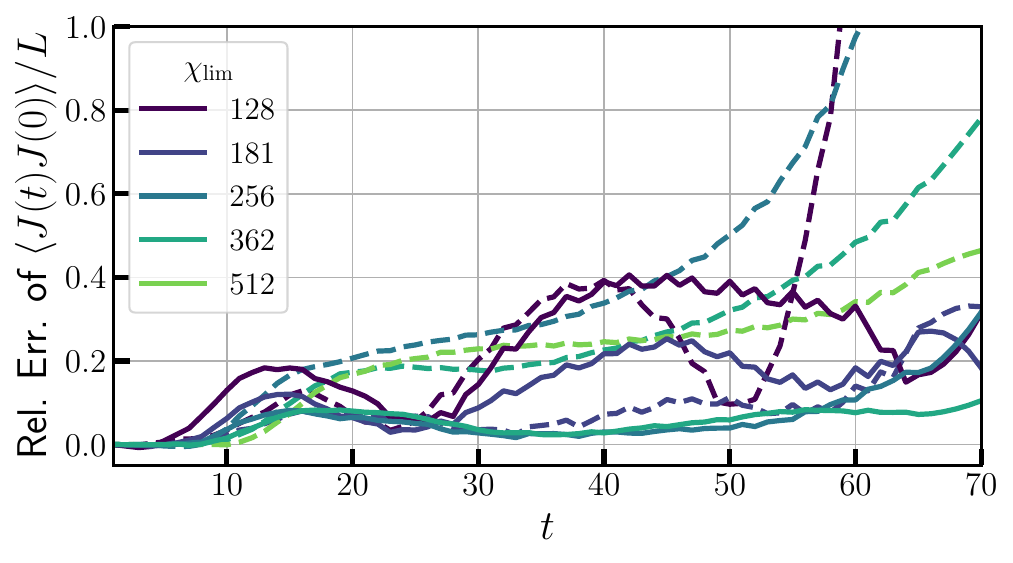}
  \caption{\label{fig:DMT-vs-TEBD} The relative error of TEBD (dashed lines) and DMT (solid lines) current correlators against the $\chi_\mathrm{lim}=512$ DMT series, smoothed with a second order Savitzky-Golay filter (computed using \texttt{scipy}~\cite{scipy}) with window width of $m = 12$.
  We smooth because low bond dimension DMT does not accurately capture the oscillatory behavior of the converged large-bond dimension simulations. This oscillatory behavior has a limited effect on the diffusion coefficient.
  We show the unsmoothed relative error in \cref{s:additional-convergence}, \cref{fig:DMT-vs-TEBD-unsmoothed}.
  }
\end{figure}

\cref{fig:dmt-tail-check} shows the current-current correlator as a function of time
as computed in DMT with time-doubling.
The correlator displays a fast decay for $t \lesssim 20$,
and an exponential decay for $20 \lesssim t \lesssim 60$
with decay rate $\approx 0.14$.
In this context, we define convergence as the relative error between $\chi_\mathrm{lim} = 256$ and $\chi_\mathrm{lim} = 512$ less than 0.15,
as shown in Fig.~\ref{fig:DMT-vs-TEBD} (also see Fig.~\ref{fig:DMT-vs-TEBD-unsmoothed}).
Because $D(t)$ is dominated by early-time contributions,
convergence in $D(t)$ (Fig.~\ref{fig:Dresults}) is much better than late-time convergence in $\expct{J(t)J(0)}$ (Fig.~\ref{fig:dmt-tail-check}).
The $t>15$ data are computed with a relatively large time step of size $\tau=0.5$ and combined with data computed using a finer time step of size $\tau=0.05$ to increase the accuracy of the numerical integral for $D(t)$.
We show convergence in Trotter step in \cref{s:additional-convergence}.

At long times, the correlator displays a curious oscillation about an overall exponential decay.
This decay and oscillation can be phenomenologically characterized using three poles as
\begin{align}
  \label{eq:phenom-pole-structure}
  \expct{J(t)J(0)} \sim A e^{-z_0t} + B e^{-z_1t} + B^* e^{-z_1^*t}\;,\ t \gtrsim 20
\end{align}
with $z_0 \approx 0.14$ a purely real decay rate and $z_1 \in \mathds C$ a decay $\Re z_1$ combined with an oscillation $\Im z_1$.
The amplitude of the oscillation changes little over the time $20 \lesssim t \lesssim 70$ so the real part of $z_1$ is fine-tuned to be close to $z_0$:
\begin{align}
  \label{eq:phenom-pole-constraint}
  |z_0 - \Re z_1| \lesssim 1/70\;.
\end{align}
Based on Fig.~\ref{fig:dmt-tail-check} we cannot rule out that $z_0 > \Re z_1$, which would lead to $\langle J(t) J(0) \rangle < 0$ at sufficiently long times.
The oscillatory behavior, together with  \eqref{eq:phenom-pole-constraint}, is consistent between the largest bond dimensions we use and across the smallest Trotter steps we use (\cref{s:additional-convergence}), and it also appears in the operator size truncated dynamics (\cref{sss:ost-results}).
We do not know if this behavior is generic, coincidental, or a result of some special feature of the mixed-field Ising model; the mechanisms causing these oscillations could be the focus of future work.

\begin{figure}[t]
\includegraphics[width=\columnwidth]{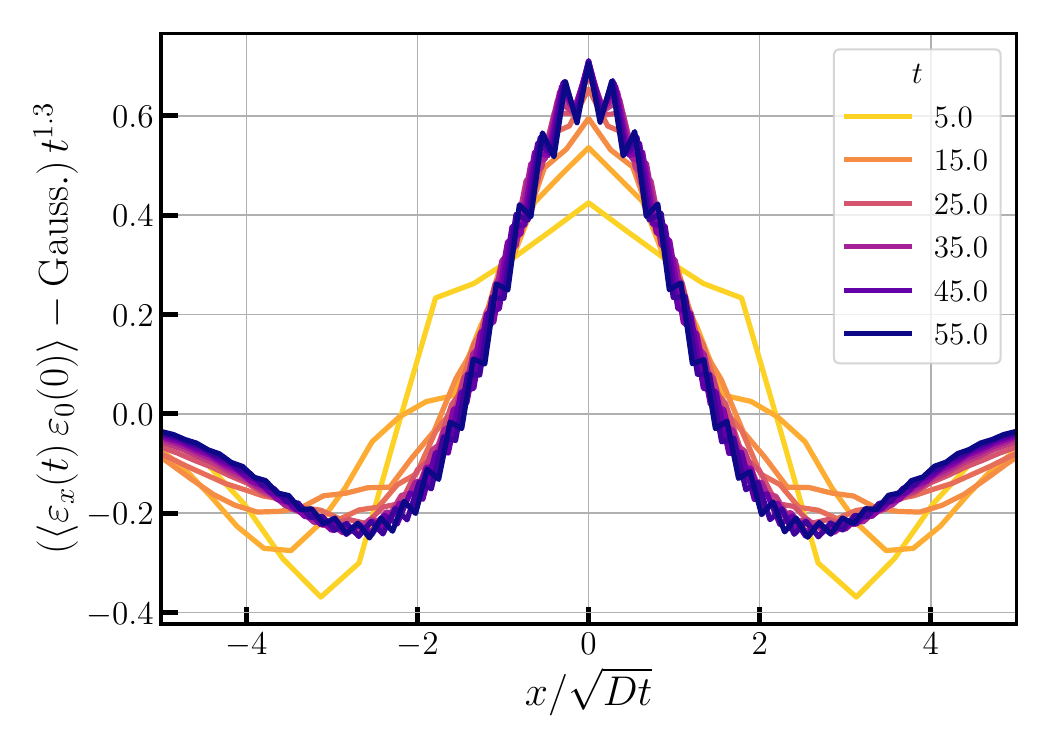}
  \caption{
  Power-law corrections to diffusion in the energy-energy correlator,  rescaled by $t^{1.3}$. The corrections are generated by subtracting a Gaussian with a variance of $2Dt$ with a diffusion constant of $D=1.446$ from the energy-energy correlator. (DMT, $\chi_\mathrm{lim}=256$, $\tau=0.5$) The scaling exponent $1.3$ yields the best approximate collapse of the corrections, as visually determined at the plotted times.
 }
  \label{fig:profile-correction}
\end{figure}

The exponential decay of the current extends over almost three decades of magnitude. 

This is in tension with the general theoretical expectation that non-linear corrections to hydrodynamics will lead to power-law decay of most hydrodynamic quantities, including the current \cite{mukerjee_statistical_2006, michailidis2023}, in the long-time limit. The inset of Fig.~\ref{fig:dmt-tail-check} shows $\expct{J(t)J(0)}$ on a log-log plot with a comparison to the theoretical prediction $\langle J(t)J(0) \rangle \sim t^{-4}$ of \onlinecite{michailidis2023} for the current decay in a system with one conserved quantity. It is evident that within the time range of our converged numerics, the current decays more rapidly than $t^{-4}$ and has no extended time regime consistent with power law decay. 
We note that this finding is consistent with the numerical results of \onlinecite{michailidis2023}, which also found decay of the current-current correlator to be faster than theoretical expectations. 
In the energy-energy correlator, however, we do see power-law corrections to diffusion, as shown in \cref{fig:profile-correction}. This is consistent with \onlinecite{michailidis2023} though the exact exponent is different, possibly due to finite time constraints.

We believe that our convergence testing (Fig.~\ref{fig:DMT-vs-TEBD}) shows true convergence, not a plateau of misleading apparent convergence, because the nonlinearities leading to long-time tails can in principle be captured by DMT.
Consider a simple description of long time tails along the lines of Mukerjee \emph{et al.}\ in \onlinecite{mukerjee_statistical_2006}.%
\footnote{
Note that Mukerjee \emph{et al.}\ \cite{mukerjee_statistical_2006} consider systems with several conserved quantities $n_\alpha$, where the leading nonlinearity was $n_\alpha \partial_x n_\beta$.
Our system, by contrast, has only one conserved quantity $\varepsilon$,
so the leading nonlinearity is $(\partial_x \varepsilon)^3$.
We have translated their argument to apply to our model,
and rephrased it in terms of operators.
}
In that description the total current $J$ is coupled to nonlinear terms like $(\partial_x \varepsilon)^3$
and each of the factors $(\partial_x \varepsilon)$ evolves according to the bare diffusive propagator.
So to see the effect of nonlinearities, the MPO must represent a three-point operator $(\partial_x\varepsilon)(\partial_y \varepsilon)(\partial_z \varepsilon)$, together with the operators required to simulate the local evolution of the three factors.
At finite time $t$, all of $x,y,z$ will be within a distance $l \sim t^{1/2}$ of each other, because each $\partial_x \varepsilon$ evolves diffusively.
Since the product of three local operators separated by $\lesssim l$ has bond dimension $\chi \propto l$, with the constant set by the complexity of the local operators,
we expect that seeing nonlinear corrections to hydrodynamics will require bond dimension
$\chi \propto l \sim t^{1/2}$.
Because the bond dimension required is polynomial in time, convergence even at moderate bond dimensions (the 256-512 shown in Fig.~\ref{fig:DMT-vs-TEBD}) is good evidence that DMT is accurately capturing even nonlinear corrections to diffusion.

\begin{figure*}[!htb]
\includegraphics[width=\textwidth]{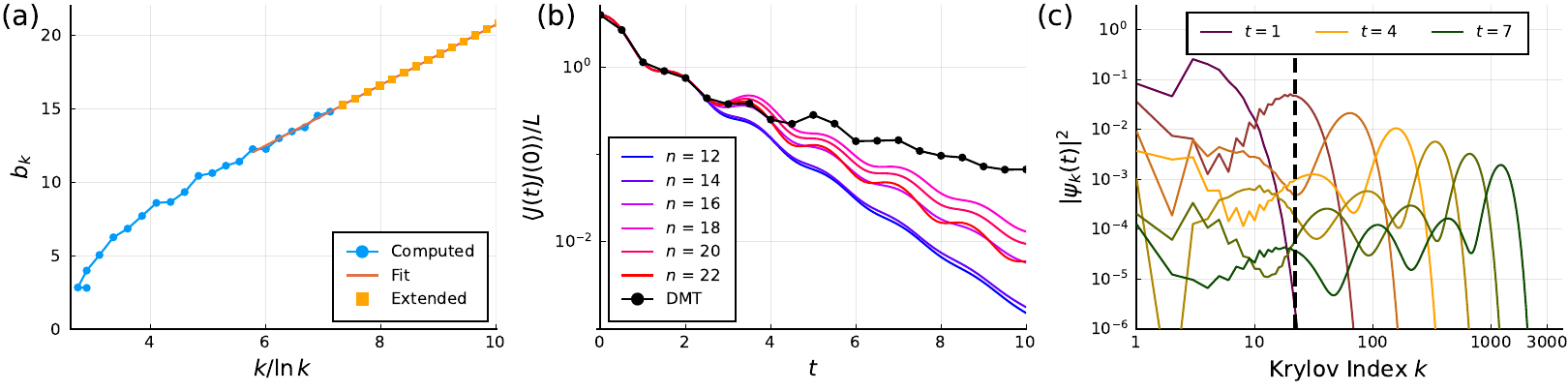}
  \caption{
    Recursion method with universal operator growth hypothesis. \textbf{(a)}: The first $22$ Krylov coefficients $b_k$ plotted against $k / \ln k$, with the linear fit used to extend the series beyond $k=22$. The fit is computed using $b_k$ for $16 \leq k \leq 22$. \textbf{(b)}: Current-current correlator $\expct{J(t)J(0)}$ computed using the extended Krylov series, using $n$ exact Krylov coefficients and extended with a fit computed using $b_k$ for $n-6 \leq k \leq n$. The results do not converge with increasing $n$, nor do they agree with the converged DMT calculations (even at short times $t \gtrsim 4$).
    \textbf{(c)}: Distribution of operator weight of $J(t)$ in the Krylov basis with $n = 22$ Krylov coefficients computed exactly. After $t_* \sim 2$ the operator has appreciable weight on the artificially extended part of the chain with $k>22$ (right of black dashed line).
 }
  \label{fig:uog-current}
\end{figure*}

We cannot, however, rule out that power-law behavior with a small coefficient dominates at times longer than those we study here,
but the amplitude of any such decay must be small compared to the exponential decay of the current that dominates on the timescales we treat.
Indeed the worsening convergence at the end of Fig.~\ref{fig:DMT-vs-TEBD}, for $t \gtrsim 60$---where we do not consider our simulations converged---might come from a transition to power-law long-time behavior that our moderate bond dimensions do not capture,
although other effects may also cause failure of convergence.

\subsection{Method comparison}

\subsubsection{TEBD against TEBD with DMT}

\cref{fig:DMT-vs-TEBD} shows how TEBD and DMT simulations converge to the $\chi_\mathrm{lim} = 512$ DMT simulations.
We measure the relative error
\begin{align}
    \left|1 - \frac{C^{JJ}_{M,\chi_\mathrm{lim}} } {C^{JJ}_{\mathrm{DMT}, 512}}\right|\;,
\end{align}
where 
\begin{equation}
    C^{JJ}_{M,\chi_\mathrm{lim}} = \frac 1 L \expct{J(t)J(0)}
\end{equation}
computed with method $M \in \{\text{TEBD, DMT}\}$ at maximum bond dimension $\chi_\mathrm{lim}$.

We find that DMT significantly outperforms TEBD at long times ($\gtrsim 50$), even at modest bond dimensions; at $t\sim 70$, the DMT results remain converged on the exponential as shown in \cref{fig:dmt-tail-check}. Though this time period contributes less to the diffusion coefficient, it is critical in determining the lack of long time tails.

At shorter times ($t\lesssim 20$), DMT performs no better than TEBD\@. Since this region overwhelmingly determines the diffusion coefficient, we do not expect greater accuracy from DMT in systems without long time tails. Contrarily, systems that do feature long-time tails may require the persisting accuracy of DMT to extract accurate diffusion coefficients.

\subsubsection{Recursion method with universal operator growth hypothesis}\label{sss:uog-results}

\cref{fig:uog-current}(b) shows the current-current correlation function $\expct{J(t)J(0)}$ computed by R-UOG
together with DMT simulations.
The R-UOG correlator agrees with DMT at short times $t \lesssim 4$,
but then departs dramatically.
That departure can be delayed only slightly by increasing $n$, the number of Krylov coefficients computed exactly.
(We stop at $n = 22$ because computing Krylov coefficients is exponentially difficult in $n$).

In \cref{fig:uog-current}(c), we plot the coefficients of  $J(t)$ in the Krylov basis, which we label $\psi_k$, as a function of time;
the dashed line marks $k = 22 $, the largest Krylov coefficient we compute exactly.
The peak of $J(t)$'s distribution crosses $k = 22 $ around $t = t_* \sim 2$.
The discrepancy between the true coefficients $b_k$ and the universal operator growth hypothesis fit can in principal affect the correlator $\expct{J(t)J(0)}$ only after the particle represented by $\psi_k(t)$ has had sufficient time to hop from $k=1$ to $k=k^*$ and back. In this simulation, this earliest possible time of divergence occurs around $t = 2t_* \sim 4$; \cref{fig:uog-current} shows that the R-UOG estimate of $\expct{J(t)J(0)}$ indeed diverges from the DMT result near this time.

We conclude that R-UOG incorrectly estimates the current-current correlator $\expct{J(t)J(0)}$ because 
the universal operator growth hypothesis fails to capture important features of the Krylov coefficients.
By design, the universal operator growth hypothesis smooths out details 
(see Fig.~\ref{fig:uog-current} left, where we plot the Krylov coefficients for our model).
The hypothesis implicitly requires that these details not affect transport coefficients.
But the departure visible in the current-current correlator of \cref{fig:uog-current}(b),
together with the amplitude distribution of \cref{fig:uog-current}(c), suggest that this is not the case.
Assuming that for larger $k^*$, the R-UOG estimate diverges from the physical value at the earliest possible time $t \sim 2 t^*$, we see that capturing the correlator to a desired time $t$ requires computing $k^* \sim \exp(t)$ Krylov coefficients accurately.

\subsubsection{Operator size truncated dynamics}\label{sss:ost-results}

The operator size truncated (OST) dynamics has two parameters:
the operator diameter $l_*$ at which we impose the absorbing boundary,
and the loss rate $\gamma$ at the absorbing boundary.
The proper loss rate can be estimated from properties of the Liouvillian graph (as in \cref{eq:gamma-semiemp}),
but it is instructive to vary $\gamma$ and probe the sensitivity of the results to that estimate.

\begin{figure}[t]
  \includegraphics[width=\columnwidth]{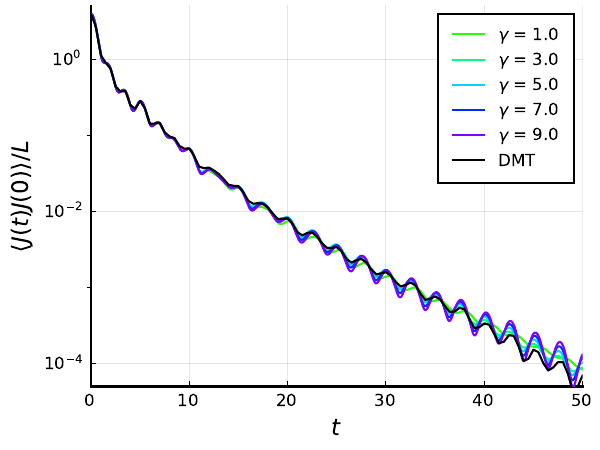}
  \caption{
  Current correlator $\expct{J(t)J(0)}$,
  computed using the operator size truncated (OST) dynamics at $l_* = 12$,
  across absorption rates $\gamma$,
  together with $\chi_\mathrm{lim} = 512$ DMT simulations.
  (The prediction \eqref{eq:gamma-semiemp} gives $\gamma = 7.01$.)
  The operator size truncated dynamics matches the overall decay of the correlator,
  but exaggerates the oscillation.
  }
  \label{fig:opgraph_vs_DMT}
\end{figure}

In \cref{fig:opgraph_vs_DMT} we show the current-current correlation function at fixed $l_* = 12$, for a variety of $\gamma$, 
together with the converged DMT calculation.
The OST dynamics accurately captures the average decay rate of the current-current correlator and the frequency of oscillations about that decay rate.
But for some $\gamma$ the oscillations appear to grow;
in the language of the phenomenological three-pole characterization 
of \cref{eq:phenom-pole-structure},
those $\gamma$ show $z_0 > \Re z_1$,
though still close to the fine-tuned
$|z_0 - \Re z_1| \lesssim 1/70$.

\begin{figure}[t]
  \includegraphics[width=\columnwidth]{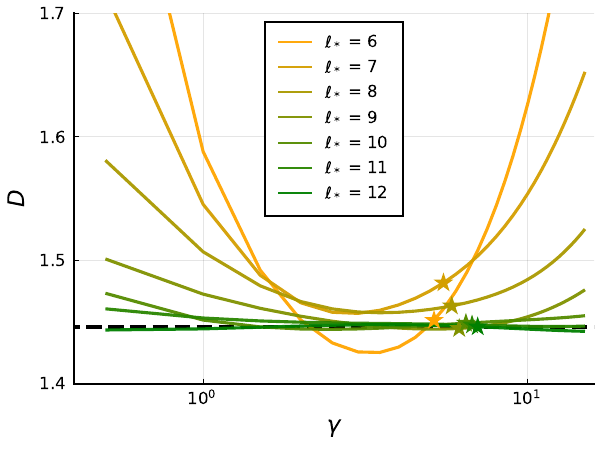}
  \caption{\label{fig:opgraph_Df_vs_gamma}
  Diffusion coefficients computed with the OST dynamics as a function of $l_*$ and loss rate $\gamma$. Stars denote the semi-empirical value of $\gamma(\ell)$   \eqref{eq:gamma-semiemp}. The dashed line marks the diffusion coefficient computed using DMT. As $l_*$ increases, the prediction becomes increasingly insensitive to the value of $\gamma$ and converges towards the DMT result.
}
\end{figure}

\cref{fig:opgraph_Df_vs_gamma} shows the predicted diffusion coefficient as a function of the loss rate $\gamma$.
For large $l_*$, the diffusion coefficient is insensitive to the loss rate $\gamma$.
This is because (in the language of \cref{eq:phenom-pole-structure})
the diffusion coefficient is controlled by the decay rate $z_0 \approx \Re z_1$,
which the OST dynamics predicts correctly.

As with DMT, one might worry that the OST dynamics artificially truncate any effects of nonlinear corrections to hydrodynamics.
As with DMT, though, we believe that for finite times and polynomially large cutoff diameters,
the OST dynamics will in fact capture those nonlinear effects,
so convergence testing in $l_*$ (and $\gamma$) is a reliable guide to the validity of our simulations.
In particular, the OST dynamics keeps arbitrarily high powers of energy density,
as long as all of the factors are close to each other (inside the cutoff diameter $l_*$).
Put another way---the OST dynamics should capture that part of the nonlinearity contained inside a diameter $l_*$ and throw away the rest.
But for times $\propto l_*^2$ this is enough,
because at finite times the relevant operators are seperated by a distance $l \propto \sqrt{t}$.
Moreover the fact that the OST dynamics agrees with the MPO dynamics,
despite truncating different portions of the nonlinearities,
gives us confidence that only the local, low-bond dimension nonlinearities captured by both methods are important on the timescales that we treat.

\section{Short-wavelength oscillatory modes: hot band second sound}\label{s:osc}

\begin{figure}[t]
  \centering
  \includegraphics[width=\columnwidth]{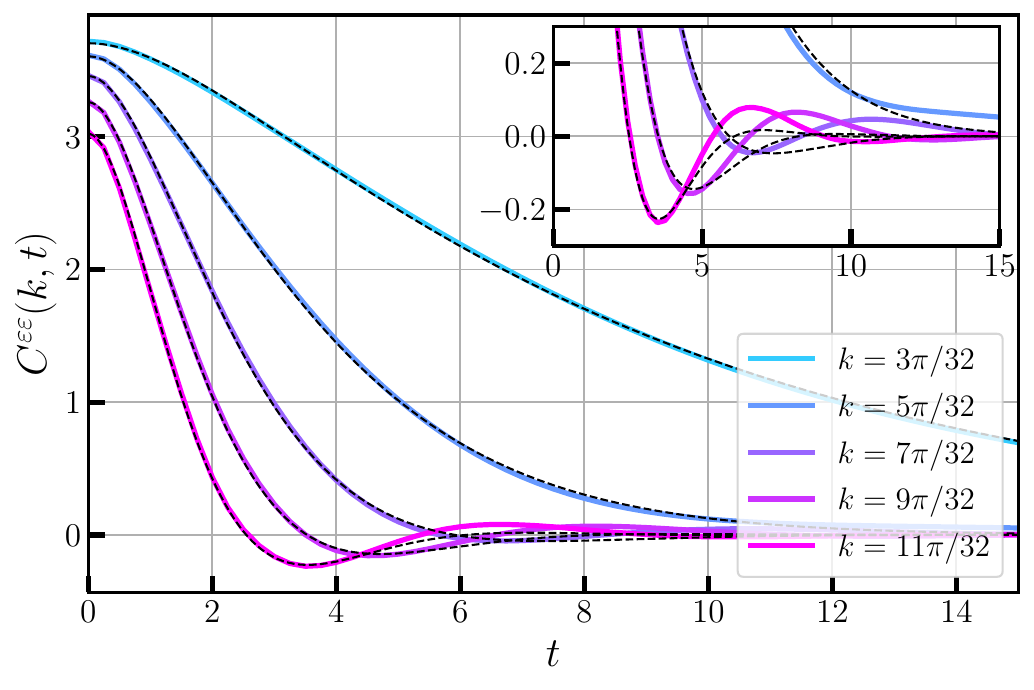}
  \centering
  \includegraphics[width=\columnwidth]{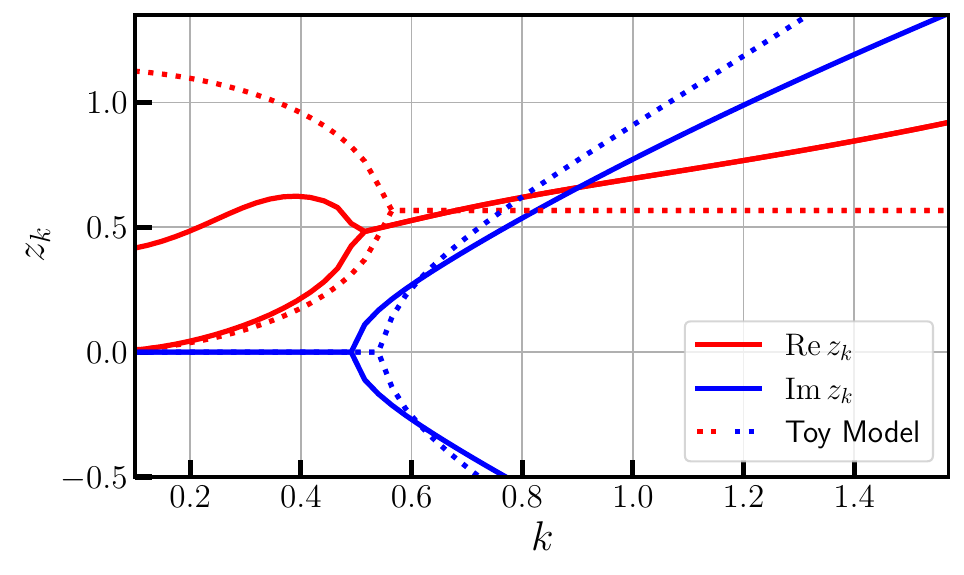}
  \centering
  \caption{
    \textbf{Top:} Energy of momentum modes over time, fit with the model in \cref{eq:fourier-fit}. Data from DMT calculation at $\chi_\mathrm{lim}=512$, $\tau=0.25$.
    \textbf{Bottom:} The complex decay rates from the fit in \cref{eq:fourier-fit}.
    Below a critical wave-vector ($k\approx 0.5$), the exponential decay follows predictable diffusion. Beyond this point, oscillations are present and the quadratic relationship disappears. Results from the toy model (\cref{eq:toy-two-modes}) are shown as otted lines.
    \label{fig:shortk}
   }
\end{figure}

\cref{fig:shortk}(top) shows the spatial Fourier transform of the energy-energy correlator 
\begin{align*}
    C^{\varepsilon\varepsilon}(k,t) = \sum_x e^{-ikx}\,\tr[\varepsilon_x(t) \varepsilon_0(0)].
\end{align*}
At small momenta, the correlator displays diffusive behavior (exponential decay with rate $D k^2$) with a diffusion coefficient $D$ that matches that computed from the integral under the zero-momentum current.

But at larger momenta, we observe underdamped oscillations of the energy-energy correlator; by analogy with~\onlinecite{bulchandaniHotBandSound2022}, we dub these oscillations hot band second sound.
The overdamped and underdamped regime can be characterized using a two-pole decay by fitting $C^{\varepsilon\varepsilon}(k,t)$ to the form
\begin{equation}
  \label{eq:fourier-fit}
  C^{\varepsilon\varepsilon}(k,t) = A_+ e^{-z_+t} + A_- e^{-z_-t}
\end{equation}
where either
\begin{align}
    z_+ = z_-^*,\quad A_+ = A_-^*
\end{align}
or
\begin{align}
    z_{\pm},A_{\pm} \in \mathds R\;.
\end{align}
We see in Fig.~\ref{fig:shortk}(top) that this fit largely matches the correlator in both the overdamped and underdamped regime for short times, though it misses some of the oscillatory detail.

Fig.~\ref{fig:shortk}(bottom) shows the fit coefficients.
At long wavelengths the decay rates are purely real $z \approx Dk^2$ with diffusion coefficient $D \approx 1.446$.
But for momenta above a sharp cutoff ($k \approx 0.50$) the Fourier modes $\varepsilon_k(t)$ become oscillatory and the decay coefficients become complex.

We can understand both the small-$k$ diffusive behavior and the oscillatory modes in terms of 
a microscopic toy model motivated by the structure of the Liouvillian graph.
In the Liouvillian graph the energy density is a state $\ket{\varepsilon_x}$ on certain diameter-1 and -2 pools;
write its Fourier transform
\begin{align}
    \ket{\varepsilon_k} = \sum_x e^{-ikx} \ket{\varepsilon_x}\;.
\end{align}
Likewise (for the Ising model \eqref{eq:ising-model}) the current operator is a state $\ket {j_x}$ on certain diameter-2 pools; 
write its Fourier transform
\begin{align}
    \ket{j_k} = \sum_x e^{-ikx} \ket{j_x}\;.
\end{align}
The continuity equation is
\begin{align}\label{eq:continuity}
  \partial_t \ket{\varepsilon_k} = 2i\sin (k/2)\ \ket{j_k}\;.
\end{align}
In terms of orthonormal operators 
$\ket{\hat \varepsilon_k} = \ket{\varepsilon_k} (\tr \varepsilon_k^\dagger \varepsilon_k)^{-1/2}$ and 
$\ket{\hat j_k} = \ket{j_k} (\tr j_k^\dagger j_k)^{-1/2}$, 
the continuity equation 
becomes a term in an effective Liouvillian
\begin{align}\label{eq:continuity-liouv}
  \sum_k 2 a_k \sin (k/2)  \ketbra{\hat \varepsilon_k}{\hat j_k} + \text{h.c.}\;
\end{align}
with
\begin{align}
    a_k^2 = \frac{\tr j_k^\dagger j_k}{\tr \varepsilon_k^\dagger \varepsilon_k}
    = \frac {2g_x^2}{1 + (g_x^2 + g_z^2) \cos^2 (k/2) } \;.
\end{align}
Note that in this section, we use the caret to denote a normalized vector as opposed to an operator. 

Additionally, weight from each current operator decays into the set of larger-diameter operators.
In the spirit of the OST effective model of \cref{ss:liouv},
mock this up by a uniform non-Hermitian decay $-i\Gamma \sum_k \ketbra{\hat j_k}{\hat j_k}$.
The toy model Liouvillian is then
\begin{align}\label{eq:liouv-ham-toy}
  \mathcal L_{\toy} &= \sum_k \Big[2 a_k \sin (k/2)  \ketbra{\hat \varepsilon_k}{\hat j_k} + \mathrm{h.c.} \notag \\
  &\qquad\qquad  -i \Gamma \ketbra{\hat j_k}{\hat j_k} \Big].
\end{align}
This Liouvillian gives an exact exponential decay $\expct{J(t)J(0)} \propto e^{-\Gamma t}$,
because the total current $J = j_{k = 0}$ is an eigenstate of $\mathcal L_\toy$ with eigenvalue $-i\Gamma$.

In terms of the memory matrix formalism 
\cite{mori_transport_1965,zwanzig_memory_1961,forster_hydrodynamic_1975}
(see \onlinecite{jungLowerBoundsConductivities2007,lucasMemoryMatrixTheory2015} for more recent discussions),
the toy model \eqref{eq:liouv-ham-toy} replaces the memory matrix $M_{\varepsilon\varepsilon}(z)$ with a single decay rate $\Gamma$.
We give an elementary, largely self-contained discussion of the memory matrix and the connection between finite momentum and $k = 0$ dynamics in \cref{ss:diffusion-spectrum}.
The toy model is also formally equivalent to the hydrodynamics with finite current relaxation rate of \onlinecite{brownBadMetallicTransport2019}.

Upon diagonalization the toy Liouvillian becomes
\begin{align}
  \mathcal L_\toy = \sum_{n\zeta} (-iz_{k\zeta}) f^\dagger_{k\zeta} f_{k\zeta}\;,
\end{align}
where $k$ is the momentum, $\zeta = \pm 1$ labels the two modes at each momentum,
$f^\dagger_{k\zeta}$ ($f_{k\zeta}$) are creation (annihilation) operators for the modes at momentum $k$, and
\begin{align}
  \label{eq:toy-two-modes}
  z_{k\zeta} = \frac \Gamma 2 \left( 1 + \zeta \sqrt{1 - \left( \frac {4a_k} \Gamma \, \sin (k/ 2)\right)^2} \right)
\end{align}
are the decay rates of the two modes.%
\footnote{At finite $k$, these decay rates are dependent on our definition of the energy density. Had we chosen a different energy density---e.g. $\varepsilon_x = 4S^z_x S^z_{x+1} + 2 g_x S^x_x + 2 g_z S^z_x$, which is not parity symmetric---we would have found a different normalization $\tr \varepsilon_k^\dagger \varepsilon_k$, hence a different $a_k$ and a different set of decay rates.
The decay rates would match at small $k$.
}
For small $k$ the $\zeta = -1$ decay rate, corresponding to the energy density, is $z_k = - (a_0^2/\Gamma) \, k^2$, allowing us to connect the diffusion coefficient to the decay rate:
\begin{align}\label{eq:gamma-D}
    \Gamma = a_0^2/D
\end{align}
We can recover the same expression by integrating the exponential decay of the current-current operator and using ~\cref{eq:current-grns2}.

\cref{fig:shortk}(bottom) shows decay rates from DMT together with the predictions of the toy model.
The toy model has a single free parameter $\Gamma$; we set $\Gamma$ using the diffusion coefficient $D=1.446$ and \eqref{eq:gamma-D}.
The toy model predicts an exceptional point---a transition from overdamped to underdamped---at $k\approx 0.55$, compared to $k \approx 0.5$ in fits to DMT data.
For $k < k_\mathrm{c}$ both modes have real decay, with $\zeta = -1$ slower;
for $k > k_\mathrm{c}$ the two modes have the same decay,
while the decay rate $z_{k\zeta}$ develops an imaginary component.
Physically this means that for long wavelength the slow modes are ``overdamped,'' and display no oscillations,
while for short wavelength the slow modes are ``underdamped,'' leading to oscillations.

The toy model quantitatively predicts the finite-wavelength dynamics
in the sense that it gives an exceptional point $k_\mathrm{c}$ within about $10\%$ of the DMT fits,
even though its only free parameter was set using behavior in the long-wavelength diffusive limit.

Although the toy model matches the behavior of the slowest-decaying mode, it also
gives a prediction for the other mode ($\zeta =+1$) that does not match the DMT data below the critical $k_\mathrm{c}$.
We speculate that the discrepancy comes from the additional, nontrivial structure in the current decay.
In the language of Sec.~\ref{ss:diffusion-spectrum}, one arrives at the toy model by replacing the complex dynamics of the current (i.e.\@ the superoperator $M_k$ in \eqref{eq:Lk-M}) by a simple exponential decay with rate $\Gamma$.
Such a single decay captures only gross, short-time features of the current-current correlator $\expct{J(t)(J(0)}$ shown in \cref{fig:dmt-tail-check,fig:opgraph_vs_DMT}.
In particular, while the toy model predicts a single decay rate, \cref{fig:dmt-tail-check,fig:opgraph_vs_DMT} shows at least two (a fast early decay for $t \lesssim 20$ and a slow long-time decay),
and the toy model does not predict the oscillations visible in \cref{fig:dmt-tail-check,fig:opgraph_vs_DMT}\;.
Moreover, the asymptotic decay rate of the correlator $\expct{J(t) J(0)}$ is slower than the value of $\Gamma$ estimated here.
The true $\zeta =+1$ decay rate may result from structure ignored by that simple approximation. %
It is not clear what particular structure results in the small-$k$ discrepancy between toy model and DMT data;
the relevant discrepancy maybe related to nonlinear corrections to the hydrodynamic description.

We do not claim that the toy model reflects the true long-time behavior of the system, since we only compare to correlators over a short time; at finite $k$, longer times become difficult to properly fit.

\section{Conclusion}
We have simulated the hydrodynamics of the fruit-fly one-dimensional (1D) nonintegrable Hamiltonian, i.e., the Ising model with tilted onsite field,
using four numerical methods:
TEBD~\cite{vidal_efficient_2003,zwolak_mixed-state_2004},
TEBD with density matrix truncation (DMT)~\cite{white2018},
the recursion method with universal operator growth hypothesis (R-UOG)~\cite{viswanathRecursionMethodApplication1994,parker2019},
and operator size truncated (OST) dynamics~\cite{whiteEffectiveDissipationRate2021}.
The universal operator growth hypothesis and the truncated Liouvillian graph methods both rely on assumptions about the dynamics of operators on the system:
R-UOG relies on an ansatz for the Lanczos coefficients of the operator dynamics
and OST relies on a chaos assumption about the dynamics of large-diameter operators.
Both assumptions require that the long-operator dynamics be unstructured.
TEBD with SVD truncation makes no such assumption---at each timestep, the SVD truncation gives a low-bond-dimension MPO that is close in Frobenius norm---but by the same token offers no guarantee that the local operators known to be important for hydrodynamics are preserved.
DMT, by contrast, imports locality into the truncation by optimizing the $L^2$ norm subject to the constraint that local operators be unchanged.
This truncation is appropriate for a wide range of systems;
it has been successfully used for thermalizing~\cite{white2018,ye_emergent_2020}
and Bethe ansatz integrable~\cite{wei_quantum_2022,ye_universal_2022} systems.
We found that DMT gives converged current-current correlators to times $t \approx 60$ with bond dimension 256,
while TEBD fails to converge for times $t \gtrsim 20$;
we also gave heuristic arguments that nonlinear hydrodynamics requires polynomial bond dimension,
putting our convergence testing on firmer conceptual ground.
We found that the converged current-current correlators show exponential decay for the times we treat;
suggesting that the coefficients of power-law long time tails---if they exist---are small enough that the tails are dominated by the exponential decay for $t \lesssim 60$.

We then used the converged DMT simulations to check the operator-size truncated (OST) dynamics and the recursion method with universal operator growth hypothesis (R-UOG).
We found that the OST dynamics matches the overall decay of the current-current correlator and gives accurate diffusion coefficients.
Indeed, although certain curious long-time oscillatory features of the correlator are sensitive to the choice of the artificial decay $\gamma$,
the diffusion coefficient is increasingly insensitive to $\gamma$ as the length scale $l^*$ is increased.
This agreement suggests that the underlying assumption of the OST dynamics---that the dynamics of long operators is chaotic and can be modelled by a simple decay---accurately reflects the portion of the dynamics that is important to transport.

The recursion method with universal operator growth hypothesis (R-UOG), by contrast, gives current-current correlators that depart from DMT simulations at $t \approx 4$,
as soon as the operator growth hypothesis becomes operative.
The R-UOG simulations also do not converge at feasible a Krylov index.
The universal operator growth hypothesis intentionally discards details of long-operator dynamics, as do all of the methods considered here.
We cannot rule out that R-UOG begins to converge for Krylov order $n \gtrsim 22$, beyond our calculations.
In particular, if we take enough Krylov coefficients to reach the regime described by the phenomenological three-pole model \eqref{eq:phenom-pole-structure},
the R-UOG may converge.
But its failure to capture the current-current correlator
(and consequently the diffusion coefficient)
at the Krylov order that we treat here suggests that the details it discards are important to transport.

We then computed a momentum-dependent energy density dynamical correlation function.
This correlation function displays a crossover from long-wavelength diffusive behavior to short-wavelength oscillatory behavior, which we dub ``hot band second sound''
by analogy with the hot band sound of~\onlinecite{bulchandaniHotBandSound2022}.
We explained this hot band second sound with a toy model rooted in the OST dynamics.
The toy model has a single free parameter, a current decay rate, which we set using the long-wavelength diffusion coefficient;
even so, it quantitatively predicts the onset of oscillations.

Our results underscore that diffusion is largely controlled by short-time behavior,
so it does not sensitively test how well a method can capture long-time dynamics.
Because the asymptotic value of the diffusion coefficient is $D = \int_0^\infty dt'\; \expct{J(t)J(0)}$,
it is dominated by early-time contributions when the current-current correlator is large.
(This is all the more true when---as we found---the current-current correlator decays exponentially.)
Consequently even methods that fail to converge at later times can still produce approximately correct values for the diffusion coefficient.
TEBD, for example, produces a diffusion coefficient within 1\% of the DMT and OST values, even though it fails to converge for $t \gtrsim 20$.
We expect similar behavior even when the diffusion coefficient is computed using some other correlator,
e.g. the energy density correlator $C^{\varepsilon\varepsilon}(x,t)$:
as long as the method gives approximately diffusive dynamics, the diffusion coefficient will be linked to the current-current correlator by a series of identities.
Reproducing a dynamical correlator such as $\expct{J(t)J(0)}$ is a more sensitive test of how well a method approximates a system's dynamics.
As DMT and the OST dynamics yield converged results for dynamical correlators over a long time interval, we expect that these methods will be useful for studying dynamics in many similar systems.

\acknowledgments{

We are grateful for helpful conversations with Brian Swingle, Mike Winer, Peter Lunts, and Alex Schuckert.
SYT thanks the Joint Quantum Institute at the University of Maryland for support through a JQI fellowship. 
This research was performed while BW held an NRC postdoctoral fellowship at the National Institute of Standards and Technology. BW was supported in part by the DoE ASCR Accelerated Research in Quantum Computing program (award No.~DE-SC0020312).
C.D.W. was supported by the U.S. Department of Energy (DOE), Office of Science, Office of Advanced Scientific Computing Research (ASCR) Quantum Computing Application Teams program, under fieldwork proposal number ERKJ347, DOE Quantum Systems Accelerator program DE-AC02-05CH11231, AFOSR MURI FA9550-22-1-0339, ARO grant W911NF-23-1-0242, ARO grant W911NF-23-1-0258, and NSF QLCI grant OMA-2120757.
JS acknowledges support from the Joint Quantum Institute. This work is also supported by the Laboratory for Physical Sciences through its continuous support of the Condensed Matter Theory Center at the University of Maryland.
}

\bibliography{bib.bib}

\begin{thebibliography}{50}%
\makeatletter
\providecommand \@ifxundefined [1]{%
 \@ifx{#1\undefined}
}%
\providecommand \@ifnum [1]{%
 \ifnum #1\expandafter \@firstoftwo
 \else \expandafter \@secondoftwo
 \fi
}%
\providecommand \@ifx [1]{%
 \ifx #1\expandafter \@firstoftwo
 \else \expandafter \@secondoftwo
 \fi
}%
\providecommand \natexlab [1]{#1}%
\providecommand \enquote  [1]{``#1''}%
\providecommand \bibnamefont  [1]{#1}%
\providecommand \bibfnamefont [1]{#1}%
\providecommand \citenamefont [1]{#1}%
\providecommand \href@noop [0]{\@secondoftwo}%
\providecommand \href [0]{\begingroup \@sanitize@url \@href}%
\providecommand \@href[1]{\@@startlink{#1}\@@href}%
\providecommand \@@href[1]{\endgroup#1\@@endlink}%
\providecommand \@sanitize@url [0]{\catcode `\\12\catcode `\$12\catcode
  `\&12\catcode `\#12\catcode `\^12\catcode `\_12\catcode `\%12\relax}%
\providecommand \@@startlink[1]{}%
\providecommand \@@endlink[0]{}%
\providecommand \url  [0]{\begingroup\@sanitize@url \@url }%
\providecommand \@url [1]{\endgroup\@href {#1}{\urlprefix }}%
\providecommand \urlprefix  [0]{URL }%
\providecommand \Eprint [0]{\href }%
\providecommand \doibase [0]{https://doi.org/}%
\providecommand \selectlanguage [0]{\@gobble}%
\providecommand \bibinfo  [0]{\@secondoftwo}%
\providecommand \bibfield  [0]{\@secondoftwo}%
\providecommand \translation [1]{[#1]}%
\providecommand \BibitemOpen [0]{}%
\providecommand \bibitemStop [0]{}%
\providecommand \bibitemNoStop [0]{.\EOS\space}%
\providecommand \EOS [0]{\spacefactor3000\relax}%
\providecommand \BibitemShut  [1]{\csname bibitem#1\endcsname}%
\let\auto@bib@innerbib\@empty
\bibitem [{\citenamefont {Brown}\ \emph {et~al.}(2019)\citenamefont {Brown},
  \citenamefont {Mitra}, \citenamefont {Guardado-Sanchez}, \citenamefont
  {Nourafkan}, \citenamefont {Reymbaut}, \citenamefont {H{\'{e}}bert},
  \citenamefont {Bergeron}, \citenamefont {Tremblay}, \citenamefont {Kokalj},
  \citenamefont {Huse}, \citenamefont {Schau{\ss}},\ and\ \citenamefont
  {Bakr}}]{brownBadMetallicTransport2019}%
  \BibitemOpen
  \bibfield  {author} {\bibinfo {author} {\bibfnamefont {P.~T.}\ \bibnamefont
  {Brown}}, \bibinfo {author} {\bibfnamefont {D.}~\bibnamefont {Mitra}},
  \bibinfo {author} {\bibfnamefont {E.}~\bibnamefont {Guardado-Sanchez}},
  \bibinfo {author} {\bibfnamefont {R.}~\bibnamefont {Nourafkan}}, \bibinfo
  {author} {\bibfnamefont {A.}~\bibnamefont {Reymbaut}}, \bibinfo {author}
  {\bibfnamefont {C.-D.}\ \bibnamefont {H{\'{e}}bert}}, \bibinfo {author}
  {\bibfnamefont {S.}~\bibnamefont {Bergeron}}, \bibinfo {author}
  {\bibfnamefont {A.-M.~S.}\ \bibnamefont {Tremblay}}, \bibinfo {author}
  {\bibfnamefont {J.}~\bibnamefont {Kokalj}}, \bibinfo {author} {\bibfnamefont
  {D.~A.}\ \bibnamefont {Huse}}, \bibinfo {author} {\bibfnamefont
  {P.}~\bibnamefont {Schau{\ss}}},\ and\ \bibinfo {author} {\bibfnamefont
  {W.~S.}\ \bibnamefont {Bakr}},\ }\bibfield  {title} {\bibinfo {title} {Bad
  metallic transport in a cold atom fermi-hubbard system},\ }\href
  {https://doi.org/10.1126/science.aat4134} {\bibfield  {journal} {\bibinfo
  {journal} {Science}\ }\textbf {\bibinfo {volume} {363}},\ \bibinfo {pages}
  {379} (\bibinfo {year} {2019})}\BibitemShut {NoStop}%
\bibitem [{\citenamefont {Wei}\ \emph {et~al.}(2022)\citenamefont {Wei},
  \citenamefont {Rubio-Abadal}, \citenamefont {Ye}, \citenamefont {Machado},
  \citenamefont {Kemp}, \citenamefont {Srakaew}, \citenamefont {Hollerith},
  \citenamefont {Rui}, \citenamefont {Gopalakrishnan}, \citenamefont {Yao},
  \citenamefont {Bloch},\ and\ \citenamefont {Zeiher}}]{wei_quantum_2022}%
  \BibitemOpen
  \bibfield  {author} {\bibinfo {author} {\bibfnamefont {D.}~\bibnamefont
  {Wei}}, \bibinfo {author} {\bibfnamefont {A.}~\bibnamefont {Rubio-Abadal}},
  \bibinfo {author} {\bibfnamefont {B.}~\bibnamefont {Ye}}, \bibinfo {author}
  {\bibfnamefont {F.}~\bibnamefont {Machado}}, \bibinfo {author} {\bibfnamefont
  {J.}~\bibnamefont {Kemp}}, \bibinfo {author} {\bibfnamefont {K.}~\bibnamefont
  {Srakaew}}, \bibinfo {author} {\bibfnamefont {S.}~\bibnamefont {Hollerith}},
  \bibinfo {author} {\bibfnamefont {J.}~\bibnamefont {Rui}}, \bibinfo {author}
  {\bibfnamefont {S.}~\bibnamefont {Gopalakrishnan}}, \bibinfo {author}
  {\bibfnamefont {N.~Y.}\ \bibnamefont {Yao}}, \bibinfo {author} {\bibfnamefont
  {I.}~\bibnamefont {Bloch}},\ and\ \bibinfo {author} {\bibfnamefont
  {J.}~\bibnamefont {Zeiher}},\ }\bibfield  {title} {\bibinfo {title} {Quantum
  gas microscopy of kardar-parisi-zhang superdiffusion},\ }\href
  {https://doi.org/10.1126/science.abk2397} {\bibfield  {journal} {\bibinfo
  {journal} {Science}\ }\textbf {\bibinfo {volume} {376}},\ \bibinfo {pages}
  {716} (\bibinfo {year} {2022})}\BibitemShut {NoStop}%
\bibitem [{\citenamefont {Zaanen}(2019)}]{zaanen2019planckian}%
  \BibitemOpen
  \bibfield  {author} {\bibinfo {author} {\bibfnamefont {J.}~\bibnamefont
  {Zaanen}},\ }\bibfield  {title} {\bibinfo {title} {Planckian dissipation,
  minimal viscosity and the transport in cuprate strange metals},\ }\href
  {https://doi.org/10.21468/SciPostPhys.6.5.061} {\bibfield  {journal}
  {\bibinfo  {journal} {SciPost Physics}\ }\textbf {\bibinfo {volume} {6}},\
  \bibinfo {pages} {061} (\bibinfo {year} {2019})}\BibitemShut {NoStop}%
\bibitem [{\citenamefont {Ayres}\ \emph {et~al.}(2021)\citenamefont {Ayres},
  \citenamefont {Berben}, \citenamefont {{\v{C}}ulo}, \citenamefont {Hsu},
  \citenamefont {van Heumen}, \citenamefont {Huang}, \citenamefont {Zaanen},
  \citenamefont {Kondo}, \citenamefont {Takeuchi}, \citenamefont {Cooper} \emph
  {et~al.}}]{ayres2021incoherent}%
  \BibitemOpen
  \bibfield  {author} {\bibinfo {author} {\bibfnamefont {J.}~\bibnamefont
  {Ayres}}, \bibinfo {author} {\bibfnamefont {M.}~\bibnamefont {Berben}},
  \bibinfo {author} {\bibfnamefont {M.}~\bibnamefont {{\v{C}}ulo}}, \bibinfo
  {author} {\bibfnamefont {Y.-T.}\ \bibnamefont {Hsu}}, \bibinfo {author}
  {\bibfnamefont {E.}~\bibnamefont {van Heumen}}, \bibinfo {author}
  {\bibfnamefont {Y.}~\bibnamefont {Huang}}, \bibinfo {author} {\bibfnamefont
  {J.}~\bibnamefont {Zaanen}}, \bibinfo {author} {\bibfnamefont
  {T.}~\bibnamefont {Kondo}}, \bibinfo {author} {\bibfnamefont
  {T.}~\bibnamefont {Takeuchi}}, \bibinfo {author} {\bibfnamefont
  {J.}~\bibnamefont {Cooper}}, \emph {et~al.},\ }\bibfield  {title} {\bibinfo
  {title} {Incoherent transport across the strange-metal regime of overdoped
  cuprates},\ }\href {https://www.nature.com/articles/s41586-021-03622-z}
  {\bibfield  {journal} {\bibinfo  {journal} {Nature}\ }\textbf {\bibinfo
  {volume} {595}},\ \bibinfo {pages} {661} (\bibinfo {year}
  {2021})}\BibitemShut {NoStop}%
\bibitem [{\citenamefont {Poniatowski}\ \emph {et~al.}(2021)\citenamefont
  {Poniatowski}, \citenamefont {Sarkar}, \citenamefont {Lobo}, \citenamefont
  {Das~Sarma},\ and\ \citenamefont {Greene}}]{poniatowski2021counterexample}%
  \BibitemOpen
  \bibfield  {author} {\bibinfo {author} {\bibfnamefont {N.~R.}\ \bibnamefont
  {Poniatowski}}, \bibinfo {author} {\bibfnamefont {T.}~\bibnamefont {Sarkar}},
  \bibinfo {author} {\bibfnamefont {R.~P. S.~M.}\ \bibnamefont {Lobo}},
  \bibinfo {author} {\bibfnamefont {S.}~\bibnamefont {Das~Sarma}},\ and\
  \bibinfo {author} {\bibfnamefont {R.~L.}\ \bibnamefont {Greene}},\ }\bibfield
   {title} {\bibinfo {title} {Counterexample to the conjectured planckian bound
  on transport},\ }\href {https://doi.org/10.1103/PhysRevB.104.235138}
  {\bibfield  {journal} {\bibinfo  {journal} {Physical Review B}\ }\textbf
  {\bibinfo {volume} {104}},\ \bibinfo {pages} {235138} (\bibinfo {year}
  {2021})}\BibitemShut {NoStop}%
\bibitem [{\citenamefont {Spivak}\ \emph {et~al.}(2010)\citenamefont {Spivak},
  \citenamefont {Kravchenko}, \citenamefont {Kivelson},\ and\ \citenamefont
  {Gao}}]{spivak2010colloquium}%
  \BibitemOpen
  \bibfield  {author} {\bibinfo {author} {\bibfnamefont {B.}~\bibnamefont
  {Spivak}}, \bibinfo {author} {\bibfnamefont {S.~V.}\ \bibnamefont
  {Kravchenko}}, \bibinfo {author} {\bibfnamefont {S.~A.}\ \bibnamefont
  {Kivelson}},\ and\ \bibinfo {author} {\bibfnamefont {X.~P.~A.}\ \bibnamefont
  {Gao}},\ }\bibfield  {title} {\bibinfo {title} {Colloquium: Transport in
  strongly correlated two dimensional electron fluids},\ }\href
  {https://doi.org/10.1103/RevModPhys.82.1743} {\bibfield  {journal} {\bibinfo
  {journal} {Reviews of modern physics}\ }\textbf {\bibinfo {volume} {82}},\
  \bibinfo {pages} {1743} (\bibinfo {year} {2010})}\BibitemShut {NoStop}%
\bibitem [{\citenamefont {Kasahara}\ \emph {et~al.}(2010)\citenamefont
  {Kasahara}, \citenamefont {Shibauchi}, \citenamefont {Hashimoto},
  \citenamefont {Ikada}, \citenamefont {Tonegawa}, \citenamefont {Okazaki},
  \citenamefont {Shishido}, \citenamefont {Ikeda}, \citenamefont {Takeya},
  \citenamefont {Hirata} \emph {et~al.}}]{kasahara2010evolution}%
  \BibitemOpen
  \bibfield  {author} {\bibinfo {author} {\bibfnamefont {S.}~\bibnamefont
  {Kasahara}}, \bibinfo {author} {\bibfnamefont {T.}~\bibnamefont {Shibauchi}},
  \bibinfo {author} {\bibfnamefont {K.}~\bibnamefont {Hashimoto}}, \bibinfo
  {author} {\bibfnamefont {K.}~\bibnamefont {Ikada}}, \bibinfo {author}
  {\bibfnamefont {S.}~\bibnamefont {Tonegawa}}, \bibinfo {author}
  {\bibfnamefont {R.}~\bibnamefont {Okazaki}}, \bibinfo {author} {\bibfnamefont
  {H.}~\bibnamefont {Shishido}}, \bibinfo {author} {\bibfnamefont
  {H.}~\bibnamefont {Ikeda}}, \bibinfo {author} {\bibfnamefont
  {H.}~\bibnamefont {Takeya}}, \bibinfo {author} {\bibfnamefont
  {K.}~\bibnamefont {Hirata}}, \emph {et~al.},\ }\bibfield  {title} {\bibinfo
  {title} {Evolution from non-fermi-to fermi-liquid transport via isovalent
  doping in bafe 2 (as 1- x p x) 2 superconductors},\ }\href
  {https://doi.org/10.1103/PhysRevB.81.184519} {\bibfield  {journal} {\bibinfo
  {journal} {Physical Review B}\ }\textbf {\bibinfo {volume} {81}},\ \bibinfo
  {pages} {184519} (\bibinfo {year} {2010})}\BibitemShut {NoStop}%
\bibitem [{\citenamefont {Sachdev}\ and\ \citenamefont
  {Keimer}(2011)}]{sachdev2011quantum}%
  \BibitemOpen
  \bibfield  {author} {\bibinfo {author} {\bibfnamefont {S.}~\bibnamefont
  {Sachdev}}\ and\ \bibinfo {author} {\bibfnamefont {B.}~\bibnamefont
  {Keimer}},\ }\bibfield  {title} {\bibinfo {title} {Quantum criticality},\
  }\href {https://doi.org/10.1063/1.3554314} {\bibfield  {journal} {\bibinfo
  {journal} {Physics Today}\ }\textbf {\bibinfo {volume} {64}},\ \bibinfo
  {pages} {29} (\bibinfo {year} {2011})}\BibitemShut {NoStop}%
\bibitem [{\citenamefont {Lucas}\ and\ \citenamefont
  {Sachdev}(2015)}]{lucasMemoryMatrixTheory2015}%
  \BibitemOpen
  \bibfield  {author} {\bibinfo {author} {\bibfnamefont {A.}~\bibnamefont
  {Lucas}}\ and\ \bibinfo {author} {\bibfnamefont {S.}~\bibnamefont
  {Sachdev}},\ }\bibfield  {title} {\bibinfo {title} {Memory matrix theory of
  magnetotransport in strange metals},\ }\href
  {https://doi.org/10.1103/PhysRevB.91.195122} {\bibfield  {journal} {\bibinfo
  {journal} {Physical Review B}\ }\textbf {\bibinfo {volume} {91}},\ \bibinfo
  {pages} {195122} (\bibinfo {year} {2015})}\BibitemShut {NoStop}%
\bibitem [{\citenamefont {Stephanov}\ \emph {et~al.}(1998)\citenamefont
  {Stephanov}, \citenamefont {Rajagopal},\ and\ \citenamefont
  {Shuryak}}]{stephanov_signatures_1998}%
  \BibitemOpen
  \bibfield  {author} {\bibinfo {author} {\bibfnamefont {M.}~\bibnamefont
  {Stephanov}}, \bibinfo {author} {\bibfnamefont {K.}~\bibnamefont
  {Rajagopal}},\ and\ \bibinfo {author} {\bibfnamefont {E.}~\bibnamefont
  {Shuryak}},\ }\bibfield  {title} {\bibinfo {title} {Signatures of the
  tricritical point in {QCD}},\ }\href
  {https://doi.org/10.1103/physrevlett.81.4816} {\bibfield  {journal} {\bibinfo
   {journal} {Physical Review Letters}\ }\textbf {\bibinfo {volume} {81}},\
  \bibinfo {pages} {4816} (\bibinfo {year} {1998})}\BibitemShut {NoStop}%
\bibitem [{\citenamefont {Kolb}\ \emph {et~al.}(2000)\citenamefont {Kolb},
  \citenamefont {Sollfrank},\ and\ \citenamefont
  {Heinz}}]{kolb_anisotropic_2000}%
  \BibitemOpen
  \bibfield  {author} {\bibinfo {author} {\bibfnamefont {P.~F.}\ \bibnamefont
  {Kolb}}, \bibinfo {author} {\bibfnamefont {J.}~\bibnamefont {Sollfrank}},\
  and\ \bibinfo {author} {\bibfnamefont {U.}~\bibnamefont {Heinz}},\ }\bibfield
   {title} {\bibinfo {title} {Anisotropic transverse flow and the quark-hadron
  phase transition},\ }\href {https://doi.org/10.1103/physrevc.62.054909}
  {\bibfield  {journal} {\bibinfo  {journal} {Physical Review C}\ }\textbf
  {\bibinfo {volume} {62}},\ \bibinfo {pages} {054909} (\bibinfo {year}
  {2000})}\BibitemShut {NoStop}%
\bibitem [{\citenamefont {Ollitrault}(1992)}]{ollitrault_anisotropy_1992}%
  \BibitemOpen
  \bibfield  {author} {\bibinfo {author} {\bibfnamefont {J.-Y.}\ \bibnamefont
  {Ollitrault}},\ }\bibfield  {title} {\bibinfo {title} {Anisotropy as a
  signature of transverse collective flow},\ }\href
  {https://doi.org/10.1103/PhysRevD.46.229} {\bibfield  {journal} {\bibinfo
  {journal} {Phys. Rev. D}\ }\textbf {\bibinfo {volume} {46}},\ \bibinfo
  {pages} {229} (\bibinfo {year} {1992})}\BibitemShut {NoStop}%
\bibitem [{\citenamefont {{STAR
  Collaboration}}(2010)}]{star_collaboration_experimental_2010}%
  \BibitemOpen
  \bibfield  {author} {\bibinfo {author} {\bibnamefont {{STAR
  Collaboration}}},\ }\bibfield  {title} {\bibinfo {title} {An {Experimental}
  {Exploration} of the {QCD} {Phase} {Diagram}: {The} {Search} for the
  {Critical} {Point} and the {Onset} of {De}-confinement},\ }\href
  {http://arxiv.org/abs/1007.2613} {\bibfield  {journal} {\bibinfo  {journal}
  {arXiv:1007.2613 [nucl-ex]}\ } (\bibinfo {year} {2010})}\BibitemShut
  {NoStop}%
\bibitem [{\citenamefont {Heinz}\ \emph {et~al.}(2015)\citenamefont {Heinz},
  \citenamefont {Sorensen}, \citenamefont {Deshpande}, \citenamefont
  {Gagliardi}, \citenamefont {Karsch}, \citenamefont {Lappi}, \citenamefont
  {Meziani}, \citenamefont {Milner}, \citenamefont {Muller}, \citenamefont
  {Nagle}, \citenamefont {Qiu}, \citenamefont {Rajagopal}, \citenamefont
  {Roland},\ and\ \citenamefont {Venugopalan}}]{heinz_exploring_2015}%
  \BibitemOpen
  \bibfield  {author} {\bibinfo {author} {\bibfnamefont {U.}~\bibnamefont
  {Heinz}}, \bibinfo {author} {\bibfnamefont {P.}~\bibnamefont {Sorensen}},
  \bibinfo {author} {\bibfnamefont {A.}~\bibnamefont {Deshpande}}, \bibinfo
  {author} {\bibfnamefont {C.}~\bibnamefont {Gagliardi}}, \bibinfo {author}
  {\bibfnamefont {F.}~\bibnamefont {Karsch}}, \bibinfo {author} {\bibfnamefont
  {T.}~\bibnamefont {Lappi}}, \bibinfo {author} {\bibfnamefont {Z.-E.}\
  \bibnamefont {Meziani}}, \bibinfo {author} {\bibfnamefont {R.}~\bibnamefont
  {Milner}}, \bibinfo {author} {\bibfnamefont {B.}~\bibnamefont {Muller}},
  \bibinfo {author} {\bibfnamefont {J.}~\bibnamefont {Nagle}}, \bibinfo
  {author} {\bibfnamefont {J.-W.}\ \bibnamefont {Qiu}}, \bibinfo {author}
  {\bibfnamefont {K.}~\bibnamefont {Rajagopal}}, \bibinfo {author}
  {\bibfnamefont {G.}~\bibnamefont {Roland}},\ and\ \bibinfo {author}
  {\bibfnamefont {R.}~\bibnamefont {Venugopalan}},\ }\bibfield  {title}
  {\bibinfo {title} {Exploring the properties of the phases of {QCD} matter -
  research opportunities and priorities for the next decade},\ }\href
  {http://arxiv.org/abs/1501.06477} {\bibfield  {journal} {\bibinfo  {journal}
  {arXiv:1501.06477 [hep-ex, physics:hep-ph, physics:nucl-ex,
  physics:nucl-th]}\ } (\bibinfo {year} {2015})}\BibitemShut {NoStop}%
\bibitem [{\citenamefont {von Keyserlingk}\ \emph {et~al.}(2022)\citenamefont
  {von Keyserlingk}, \citenamefont {Pollmann},\ and\ \citenamefont
  {Rakovszky}}]{von_keyserlingk_operator_2021}%
  \BibitemOpen
  \bibfield  {author} {\bibinfo {author} {\bibfnamefont {C.}~\bibnamefont {von
  Keyserlingk}}, \bibinfo {author} {\bibfnamefont {F.}~\bibnamefont
  {Pollmann}},\ and\ \bibinfo {author} {\bibfnamefont {T.}~\bibnamefont
  {Rakovszky}},\ }\bibfield  {title} {\bibinfo {title} {Operator backflow and
  the classical simulation of quantum transport},\ }\href
  {https://doi.org/10.1103/physrevb.105.245101} {\bibfield  {journal} {\bibinfo
   {journal} {Physical Review B}\ }\textbf {\bibinfo {volume} {105}},\ \bibinfo
  {pages} {245101} (\bibinfo {year} {2022})}\BibitemShut {NoStop}%
\bibitem [{\citenamefont {White}\ \emph {et~al.}(2018)\citenamefont {White},
  \citenamefont {Zaletel}, \citenamefont {Mong},\ and\ \citenamefont
  {Refael}}]{white2018}%
  \BibitemOpen
  \bibfield  {author} {\bibinfo {author} {\bibfnamefont {C.~D.}\ \bibnamefont
  {White}}, \bibinfo {author} {\bibfnamefont {M.}~\bibnamefont {Zaletel}},
  \bibinfo {author} {\bibfnamefont {R.~S.~K.}\ \bibnamefont {Mong}},\ and\
  \bibinfo {author} {\bibfnamefont {G.}~\bibnamefont {Refael}},\ }\bibfield
  {title} {\bibinfo {title} {Quantum dynamics of thermalizing systems},\ }\href
  {https://doi.org/10.1103/physrevb.97.035127} {\bibfield  {journal} {\bibinfo
  {journal} {Physical Review B}\ }\textbf {\bibinfo {volume} {97}},\ \bibinfo
  {pages} {035127} (\bibinfo {year} {2018})}\BibitemShut {NoStop}%
\bibitem [{\citenamefont {Rakovszky}\ \emph {et~al.}(2022)\citenamefont
  {Rakovszky}, \citenamefont {von Keyserlingk},\ and\ \citenamefont
  {Pollmann}}]{rakovszky2022}%
  \BibitemOpen
  \bibfield  {author} {\bibinfo {author} {\bibfnamefont {T.}~\bibnamefont
  {Rakovszky}}, \bibinfo {author} {\bibfnamefont {C.~W.}\ \bibnamefont {von
  Keyserlingk}},\ and\ \bibinfo {author} {\bibfnamefont {F.}~\bibnamefont
  {Pollmann}},\ }\bibfield  {title} {\bibinfo {title} {Dissipation-assisted
  operator evolution method for capturing hydrodynamic transport},\ }\href
  {https://doi.org/10.1103/physrevb.105.075131} {\bibfield  {journal} {\bibinfo
   {journal} {Physical Review B}\ }\textbf {\bibinfo {volume} {105}},\ \bibinfo
  {pages} {075131} (\bibinfo {year} {2022})}\BibitemShut {NoStop}%
\bibitem [{\citenamefont {Kvorning}\ \emph {et~al.}(2022)\citenamefont
  {Kvorning}, \citenamefont {Herviou},\ and\ \citenamefont
  {Bardarson}}]{kvorningTimeevolutionLocalInformation2021}%
  \BibitemOpen
  \bibfield  {author} {\bibinfo {author} {\bibfnamefont {T.~K.}\ \bibnamefont
  {Kvorning}}, \bibinfo {author} {\bibfnamefont {L.}~\bibnamefont {Herviou}},\
  and\ \bibinfo {author} {\bibfnamefont {J.~H.}\ \bibnamefont {Bardarson}},\
  }\bibfield  {title} {\bibinfo {title} {{Time-evolution of local information:
  thermalization dynamics of local observables}},\ }\href
  {https://doi.org/10.21468/SciPostPhys.13.4.080} {\bibfield  {journal}
  {\bibinfo  {journal} {SciPost Phys.}\ }\textbf {\bibinfo {volume} {13}},\
  \bibinfo {pages} {080} (\bibinfo {year} {2022})}\BibitemShut {NoStop}%
\bibitem [{\citenamefont {Parker}\ \emph {et~al.}(2019)\citenamefont {Parker},
  \citenamefont {Cao}, \citenamefont {Avdoshkin}, \citenamefont {Scaffidi},\
  and\ \citenamefont {Altman}}]{parker2019}%
  \BibitemOpen
  \bibfield  {author} {\bibinfo {author} {\bibfnamefont {D.~E.}\ \bibnamefont
  {Parker}}, \bibinfo {author} {\bibfnamefont {X.}~\bibnamefont {Cao}},
  \bibinfo {author} {\bibfnamefont {A.}~\bibnamefont {Avdoshkin}}, \bibinfo
  {author} {\bibfnamefont {T.}~\bibnamefont {Scaffidi}},\ and\ \bibinfo
  {author} {\bibfnamefont {E.}~\bibnamefont {Altman}},\ }\bibfield  {title}
  {\bibinfo {title} {A universal operator growth hypothesis},\ }\href
  {https://doi.org/10.1103/physrevx.9.041017} {\bibfield  {journal} {\bibinfo
  {journal} {Physical Review X}\ }\textbf {\bibinfo {volume} {9}},\ \bibinfo
  {pages} {041017} (\bibinfo {year} {2019})}\BibitemShut {NoStop}%
\bibitem [{\citenamefont {White}(2023)}]{whiteEffectiveDissipationRate2021}%
  \BibitemOpen
  \bibfield  {author} {\bibinfo {author} {\bibfnamefont {C.~D.}\ \bibnamefont
  {White}},\ }\bibfield  {title} {\bibinfo {title} {Effective dissipation rate
  in a liouvillian-graph picture of high-temperature quantum hydrodynamics},\
  }\href {https://doi.org/10.1103/physrevb.107.094311} {\bibfield  {journal}
  {\bibinfo  {journal} {Physical Review B}\ }\textbf {\bibinfo {volume}
  {107}},\ \bibinfo {pages} {094311} (\bibinfo {year} {2023})}\BibitemShut
  {NoStop}%
\bibitem [{\citenamefont {Vidal}(2003)}]{vidal_efficient_2003}%
  \BibitemOpen
  \bibfield  {author} {\bibinfo {author} {\bibfnamefont {G.}~\bibnamefont
  {Vidal}},\ }\bibfield  {title} {\bibinfo {title} {Efficient classical
  simulation of slightly entangled quantum computations},\ }\href
  {https://doi.org/10.1103/physrevlett.91.147902} {\bibfield  {journal}
  {\bibinfo  {journal} {Physical Review Letters}\ }\textbf {\bibinfo {volume}
  {91}},\ \bibinfo {pages} {147902} (\bibinfo {year} {2003})}\BibitemShut
  {NoStop}%
\bibitem [{\citenamefont {Vidal}(2004)}]{vidal_efficient_2004}%
  \BibitemOpen
  \bibfield  {author} {\bibinfo {author} {\bibfnamefont {G.}~\bibnamefont
  {Vidal}},\ }\bibfield  {title} {\bibinfo {title} {Efficient simulation of
  one-dimensional quantum many-body systems},\ }\href
  {https://doi.org/10.1103/physrevlett.93.040502} {\bibfield  {journal}
  {\bibinfo  {journal} {Physical Review Letters}\ }\textbf {\bibinfo {volume}
  {93}},\ \bibinfo {pages} {040502} (\bibinfo {year} {2004})}\BibitemShut
  {NoStop}%
\bibitem [{\citenamefont {Zwolak}\ and\ \citenamefont
  {Vidal}(2004)}]{zwolak_mixed-state_2004}%
  \BibitemOpen
  \bibfield  {author} {\bibinfo {author} {\bibfnamefont {M.}~\bibnamefont
  {Zwolak}}\ and\ \bibinfo {author} {\bibfnamefont {G.}~\bibnamefont {Vidal}},\
  }\bibfield  {title} {\bibinfo {title} {Mixed-state dynamics in
  one-dimensional quantum lattice systems: A time-dependent superoperator
  renormalization algorithm},\ }\href
  {https://doi.org/10.1103/physrevlett.93.207205} {\bibfield  {journal}
  {\bibinfo  {journal} {Physical Review Letters}\ }\textbf {\bibinfo {volume}
  {93}},\ \bibinfo {pages} {207205} (\bibinfo {year} {2004})}\BibitemShut
  {NoStop}%
\bibitem [{\citenamefont {Leviatan}\ \emph {et~al.}(2017)\citenamefont
  {Leviatan}, \citenamefont {Pollmann}, \citenamefont {Bardarson},
  \citenamefont {Huse},\ and\ \citenamefont
  {Altman}}]{leviatanQuantumThermalizationDynamics2017}%
  \BibitemOpen
  \bibfield  {author} {\bibinfo {author} {\bibfnamefont {E.}~\bibnamefont
  {Leviatan}}, \bibinfo {author} {\bibfnamefont {F.}~\bibnamefont {Pollmann}},
  \bibinfo {author} {\bibfnamefont {J.~H.}\ \bibnamefont {Bardarson}}, \bibinfo
  {author} {\bibfnamefont {D.~A.}\ \bibnamefont {Huse}},\ and\ \bibinfo
  {author} {\bibfnamefont {E.}~\bibnamefont {Altman}},\ }\bibfield  {title}
  {\bibinfo {title} {Quantum thermalization dynamics with {{Matrix-Product
  States}}},\ }\href {http://arxiv.org/abs/1702.08894} {\bibfield  {journal}
  {\bibinfo  {journal} {arXiv:1702.08894 [cond-mat, physics:quant-ph]}\ }
  (\bibinfo {year} {2017})},\ \Eprint {https://arxiv.org/abs/1702.08894}
  {arxiv:1702.08894 [cond-mat, physics:quant-ph]} \BibitemShut {NoStop}%
\bibitem [{\citenamefont {Bulchandani}\ and\ \citenamefont
  {Huse}(2022)}]{bulchandaniHotBandSound2022}%
  \BibitemOpen
  \bibfield  {author} {\bibinfo {author} {\bibfnamefont {V.~B.}\ \bibnamefont
  {Bulchandani}}\ and\ \bibinfo {author} {\bibfnamefont {D.~A.}\ \bibnamefont
  {Huse}},\ }\href {https://doi.org/10.48550/ARXIV.2208.13767} {\bibinfo
  {title} {Hot band sound}} (\bibinfo {year} {2022}),\ \Eprint
  {https://arxiv.org/abs/2208.13767} {arXiv:2208.13767} \BibitemShut {NoStop}%
\bibitem [{\citenamefont {Kim}\ and\ \citenamefont
  {Huse}(2013)}]{kim_ballistic_2013}%
  \BibitemOpen
  \bibfield  {author} {\bibinfo {author} {\bibfnamefont {H.}~\bibnamefont
  {Kim}}\ and\ \bibinfo {author} {\bibfnamefont {D.~A.}\ \bibnamefont {Huse}},\
  }\bibfield  {title} {\bibinfo {title} {Ballistic spreading of entanglement in
  a diffusive nonintegrable system},\ }\href
  {https://doi.org/10.1103/physrevlett.111.127205} {\bibfield  {journal}
  {\bibinfo  {journal} {Physical Review Letters}\ }\textbf {\bibinfo {volume}
  {111}},\ \bibinfo {pages} {127205} (\bibinfo {year} {2013})}\BibitemShut
  {NoStop}%
\bibitem [{\citenamefont {Kim}\ \emph {et~al.}(2014)\citenamefont {Kim},
  \citenamefont {Ikeda},\ and\ \citenamefont {Huse}}]{kim_testing_2014}%
  \BibitemOpen
  \bibfield  {author} {\bibinfo {author} {\bibfnamefont {H.}~\bibnamefont
  {Kim}}, \bibinfo {author} {\bibfnamefont {T.~N.}\ \bibnamefont {Ikeda}},\
  and\ \bibinfo {author} {\bibfnamefont {D.~A.}\ \bibnamefont {Huse}},\
  }\bibfield  {title} {\bibinfo {title} {Testing whether all eigenstates obey
  the eigenstate thermalization hypothesis},\ }\href
  {https://doi.org/10.1103/physreve.90.052105} {\bibfield  {journal} {\bibinfo
  {journal} {Physical Review E}\ }\textbf {\bibinfo {volume} {90}},\ \bibinfo
  {pages} {052105} (\bibinfo {year} {2014})}\BibitemShut {NoStop}%
\bibitem [{\citenamefont {Landau}\ and\ \citenamefont
  {Lifshitz}(2013)}]{landau2013fluid}%
  \BibitemOpen
  \bibfield  {author} {\bibinfo {author} {\bibfnamefont {L.~D.}\ \bibnamefont
  {Landau}}\ and\ \bibinfo {author} {\bibfnamefont {E.~M.}\ \bibnamefont
  {Lifshitz}},\ }\href@noop {} {\emph {\bibinfo {title} {Fluid Mechanics:
  Landau and Lifshitz: Course of Theoretical Physics, Volume 6}}},\
  Vol.~\bibinfo {volume} {6}\ (\bibinfo  {publisher} {Elsevier},\ \bibinfo
  {year} {2013})\BibitemShut {NoStop}%
\bibitem [{\citenamefont {Doyon}(2020)}]{doyonLectureNotesGeneralised2020}%
  \BibitemOpen
  \bibfield  {author} {\bibinfo {author} {\bibfnamefont {B.}~\bibnamefont
  {Doyon}},\ }\bibfield  {title} {\bibinfo {title} {Lecture notes on
  {{Generalised Hydrodynamics}}},\ }\href
  {https://doi.org/10.21468/SciPostPhysLectNotes.18} {\bibfield  {journal}
  {\bibinfo  {journal} {SciPost Physics Lecture Notes}\ ,\ \bibinfo {pages}
  {018}} (\bibinfo {year} {2020})}\BibitemShut {NoStop}%
\bibitem [{\citenamefont {Nardis}\ and\ \citenamefont
  {Doyon}(2023)}]{nardisHydrodynamicGaugeFixing2023}%
  \BibitemOpen
  \bibfield  {author} {\bibinfo {author} {\bibfnamefont {J.~D.}\ \bibnamefont
  {Nardis}}\ and\ \bibinfo {author} {\bibfnamefont {B.}~\bibnamefont {Doyon}},\
  }\bibfield  {title} {\bibinfo {title} {Hydrodynamic gauge fixing and higher
  order hydrodynamic expansion},\ }\href
  {https://doi.org/10.1088/1751-8121/acd153} {\bibfield  {journal} {\bibinfo
  {journal} {Journal of Physics A: Mathematical and Theoretical}\ }\textbf
  {\bibinfo {volume} {56}},\ \bibinfo {pages} {245001} (\bibinfo {year}
  {2023})}\BibitemShut {NoStop}%
\bibitem [{\citenamefont {Steinigeweg}\ \emph {et~al.}(2009)\citenamefont
  {Steinigeweg}, \citenamefont {Wichterich},\ and\ \citenamefont
  {Gemmer}}]{steinigewegDensityDynamicsCurrent2009}%
  \BibitemOpen
  \bibfield  {author} {\bibinfo {author} {\bibfnamefont {R.}~\bibnamefont
  {Steinigeweg}}, \bibinfo {author} {\bibfnamefont {H.}~\bibnamefont
  {Wichterich}},\ and\ \bibinfo {author} {\bibfnamefont {J.}~\bibnamefont
  {Gemmer}},\ }\bibfield  {title} {\bibinfo {title} {Density dynamics from
  current auto-correlations at finite time- and length-scales},\ }\href
  {https://doi.org/10.1209/0295-5075/88/10004} {\bibfield  {journal} {\bibinfo
  {journal} {{EPL} (Europhysics Letters)}\ }\textbf {\bibinfo {volume} {88}},\
  \bibinfo {pages} {10004} (\bibinfo {year} {2009})}\BibitemShut {NoStop}%
\bibitem [{\citenamefont {Mukerjee}\ \emph {et~al.}(2006)\citenamefont
  {Mukerjee}, \citenamefont {Oganesyan},\ and\ \citenamefont
  {Huse}}]{mukerjee_statistical_2006}%
  \BibitemOpen
  \bibfield  {author} {\bibinfo {author} {\bibfnamefont {S.}~\bibnamefont
  {Mukerjee}}, \bibinfo {author} {\bibfnamefont {V.}~\bibnamefont
  {Oganesyan}},\ and\ \bibinfo {author} {\bibfnamefont {D.}~\bibnamefont
  {Huse}},\ }\bibfield  {title} {\bibinfo {title} {Statistical theory of
  transport by strongly interacting lattice fermions},\ }\href
  {https://doi.org/10.1103/physrevb.73.035113} {\bibfield  {journal} {\bibinfo
  {journal} {Physical Review B}\ }\textbf {\bibinfo {volume} {73}},\ \bibinfo
  {pages} {035113} (\bibinfo {year} {2006})}\BibitemShut {NoStop}%
\bibitem [{\citenamefont {Forster}(1975)}]{forster_hydrodynamic_1975}%
  \BibitemOpen
  \bibfield  {author} {\bibinfo {author} {\bibfnamefont {D.}~\bibnamefont
  {Forster}},\ }\href@noop {} {\emph {\bibinfo {title} {Hydrodynamic
  fluctuations, broken symmetry, and correlation functions}}},\ Frontiers in
  physics ; 47\ (\bibinfo  {publisher} {W.A. Benjamin, Advanced Book Program},\
  \bibinfo {address} {Reading, Mass.},\ \bibinfo {year} {1975})\BibitemShut
  {NoStop}%
\bibitem [{\citenamefont {Hild}\ \emph {et~al.}(2014)\citenamefont {Hild},
  \citenamefont {Fukuhara}, \citenamefont {Schauss}, \citenamefont {Zeiher},
  \citenamefont {Knap}, \citenamefont {Demler}, \citenamefont {Bloch},\ and\
  \citenamefont {Gross}}]{Hild2014}%
  \BibitemOpen
  \bibfield  {author} {\bibinfo {author} {\bibfnamefont {S.}~\bibnamefont
  {Hild}}, \bibinfo {author} {\bibfnamefont {T.}~\bibnamefont {Fukuhara}},
  \bibinfo {author} {\bibfnamefont {P.}~\bibnamefont {Schauss}}, \bibinfo
  {author} {\bibfnamefont {J.}~\bibnamefont {Zeiher}}, \bibinfo {author}
  {\bibfnamefont {M.}~\bibnamefont {Knap}}, \bibinfo {author} {\bibfnamefont
  {E.}~\bibnamefont {Demler}}, \bibinfo {author} {\bibfnamefont
  {I.}~\bibnamefont {Bloch}},\ and\ \bibinfo {author} {\bibfnamefont
  {C.}~\bibnamefont {Gross}},\ }\bibfield  {title} {\bibinfo {title}
  {Far-from-equilibrium spin transport in heisenberg quantum magnets},\ }\href
  {https://doi.org/10.1103/physrevlett.113.147205} {\bibfield  {journal}
  {\bibinfo  {journal} {Physical Review Letters}\ }\textbf {\bibinfo {volume}
  {113}},\ \bibinfo {pages} {147205} (\bibinfo {year} {2014})}\BibitemShut
  {NoStop}%
\bibitem [{\citenamefont
  {Schollw{\"{o}}ck}(2011)}]{schollwoeck_density-matrix_2011}%
  \BibitemOpen
  \bibfield  {author} {\bibinfo {author} {\bibfnamefont {U.}~\bibnamefont
  {Schollw{\"{o}}ck}},\ }\bibfield  {title} {\bibinfo {title} {The
  density-matrix renormalization group in the age of matrix product states},\
  }\href {https://doi.org/10.1016/j.aop.2010.09.012} {\bibfield  {journal}
  {\bibinfo  {journal} {Annals of Physics}\ }\textbf {\bibinfo {volume}
  {326}},\ \bibinfo {pages} {96} (\bibinfo {year} {2011})}\BibitemShut
  {NoStop}%
\bibitem [{\citenamefont {Barthel}\ and\ \citenamefont
  {Zhang}(2020)}]{barthel2020}%
  \BibitemOpen
  \bibfield  {author} {\bibinfo {author} {\bibfnamefont {T.}~\bibnamefont
  {Barthel}}\ and\ \bibinfo {author} {\bibfnamefont {Y.}~\bibnamefont
  {Zhang}},\ }\bibfield  {title} {\bibinfo {title} {Optimized
  lie{\textendash}trotter{\textendash}suzuki decompositions for two and three
  non-commuting terms},\ }\href {https://doi.org/10.1016/j.aop.2020.168165}
  {\bibfield  {journal} {\bibinfo  {journal} {Annals of Physics}\ }\textbf
  {\bibinfo {volume} {418}},\ \bibinfo {pages} {168165} (\bibinfo {year}
  {2020})}\BibitemShut {NoStop}%
\bibitem [{\citenamefont {Mattis}(1981)}]{mattisHowReducePractically1981}%
  \BibitemOpen
  \bibfield  {author} {\bibinfo {author} {\bibfnamefont {D.~C.}\ \bibnamefont
  {Mattis}},\ }\bibfield  {title} {\bibinfo {title} {How to reduce practically
  any problem to one dimension},\ }in\ \href
  {https://doi.org/10.1007/978-3-642-81592-8_1} {\emph {\bibinfo {booktitle}
  {Springer Series in Solid-State Sciences}}}\ (\bibinfo  {publisher} {Springer
  Berlin Heidelberg},\ \bibinfo {year} {1981})\ pp.\ \bibinfo {pages}
  {3--10}\BibitemShut {NoStop}%
\bibitem [{\citenamefont {Viswanath}\ and\ \citenamefont
  {M{\"{u}}ller}(1994)}]{viswanathRecursionMethodApplication1994}%
  \BibitemOpen
  \bibfield  {author} {\bibinfo {author} {\bibfnamefont {V.~S.}\ \bibnamefont
  {Viswanath}}\ and\ \bibinfo {author} {\bibfnamefont {G.}~\bibnamefont
  {M{\"{u}}ller}},\ }\href {https://doi.org/10.1007/978-3-540-48651-0} {\emph
  {\bibinfo {title} {The Recursion Method}}}\ (\bibinfo  {publisher} {Springer
  Berlin Heidelberg},\ \bibinfo {year} {1994})\BibitemShut {NoStop}%
\bibitem [{\citenamefont {Altman}(2018)}]{altman_ehud_computing_2018}%
  \BibitemOpen
  \bibfield  {author} {\bibinfo {author} {\bibfnamefont {E.}~\bibnamefont
  {Altman}},\ }\href {https://online.kitp.ucsb.edu/online/dynq_c18/altman/}
  {\bibinfo {title} {Computing quantum thermalization dynamics}} (\bibinfo
  {year} {2018})\BibitemShut {NoStop}%
\bibitem [{\citenamefont {Ehrenfest}\ and\ \citenamefont
  {Ehrenfest}(2002)}]{ehrenfest_conceptual_2002}%
  \BibitemOpen
  \bibfield  {author} {\bibinfo {author} {\bibfnamefont {P.}~\bibnamefont
  {Ehrenfest}}\ and\ \bibinfo {author} {\bibfnamefont {T.}~\bibnamefont
  {Ehrenfest}},\ }\href@noop {} {\emph {\bibinfo {title} {The {Conceptual}
  {Foundations} of the {Statistical} {Approach} in {Mechanics}}}}\ (\bibinfo
  {publisher} {Dover Publications},\ \bibinfo {year} {2002})\BibitemShut
  {NoStop}%
\bibitem [{\citenamefont {Gottwald}\ and\ \citenamefont
  {Oliver}(2009)}]{gottwald_boltzmanns_2009}%
  \BibitemOpen
  \bibfield  {author} {\bibinfo {author} {\bibfnamefont {G.~A.}\ \bibnamefont
  {Gottwald}}\ and\ \bibinfo {author} {\bibfnamefont {M.}~\bibnamefont
  {Oliver}},\ }\bibfield  {title} {\bibinfo {title}
  {Boltzmann{\textquotesingle}s dilemma: An introduction to statistical
  mechanics via the kac ring},\ }\href {https://doi.org/10.1137/070705799}
  {\bibfield  {journal} {\bibinfo  {journal} {{SIAM} Review}\ }\textbf
  {\bibinfo {volume} {51}},\ \bibinfo {pages} {613} (\bibinfo {year}
  {2009})}\BibitemShut {NoStop}%
\bibitem [{\citenamefont {Zwanzig}(1961)}]{zwanzig_memory_1961}%
  \BibitemOpen
  \bibfield  {author} {\bibinfo {author} {\bibfnamefont {R.}~\bibnamefont
  {Zwanzig}},\ }\bibfield  {title} {\bibinfo {title} {Memory effects in
  irreversible thermodynamics},\ }\href
  {https://doi.org/10.1103/physrev.124.983} {\bibfield  {journal} {\bibinfo
  {journal} {Physical Review}\ }\textbf {\bibinfo {volume} {124}},\ \bibinfo
  {pages} {983} (\bibinfo {year} {1961})}\BibitemShut {NoStop}%
\bibitem [{\citenamefont {Mori}(1965)}]{mori_transport_1965}%
  \BibitemOpen
  \bibfield  {author} {\bibinfo {author} {\bibfnamefont {H.}~\bibnamefont
  {Mori}},\ }\bibfield  {title} {\bibinfo {title} {Transport, collective
  motion, and brownian motion},\ }\href {https://doi.org/10.1143/ptp.33.423}
  {\bibfield  {journal} {\bibinfo  {journal} {Progress of Theoretical Physics}\
  }\textbf {\bibinfo {volume} {33}},\ \bibinfo {pages} {423} (\bibinfo {year}
  {1965})}\BibitemShut {NoStop}%
\bibitem [{\citenamefont {Haegeman}(2024)}]{juthoKrylovKitJl2023}%
  \BibitemOpen
  \bibfield  {author} {\bibinfo {author} {\bibfnamefont {J.}~\bibnamefont
  {Haegeman}},\ }\href {https://doi.org/10.5281/zenodo.10884302} {\bibinfo
  {title} {Krylovkit}} (\bibinfo {year} {2024})\BibitemShut {NoStop}%
\bibitem [{\citenamefont {Michailidis}\ \emph {et~al.}(2024)\citenamefont
  {Michailidis}, \citenamefont {Abanin},\ and\ \citenamefont
  {Delacr\'etaz}}]{michailidis2023}%
  \BibitemOpen
  \bibfield  {author} {\bibinfo {author} {\bibfnamefont {A.~A.}\ \bibnamefont
  {Michailidis}}, \bibinfo {author} {\bibfnamefont {D.~A.}\ \bibnamefont
  {Abanin}},\ and\ \bibinfo {author} {\bibfnamefont {L.~V.}\ \bibnamefont
  {Delacr\'etaz}},\ }\bibfield  {title} {\bibinfo {title} {Corrections to
  diffusion in interacting quantum systems},\ }\href
  {https://doi.org/10.1103/PhysRevX.14.031020} {\bibfield  {journal} {\bibinfo
  {journal} {Phys. Rev. X}\ }\textbf {\bibinfo {volume} {14}},\ \bibinfo
  {pages} {031020} (\bibinfo {year} {2024})}\BibitemShut {NoStop}%
\bibitem [{\citenamefont {Virtanen}\ \emph {et~al.}(2020)\citenamefont
  {Virtanen}, \citenamefont {Gommers}, \citenamefont {Oliphant}, \citenamefont
  {Haberland}, \citenamefont {Reddy}, \citenamefont {Cournapeau}, \citenamefont
  {Burovski}, \citenamefont {Peterson}, \citenamefont {Weckesser},
  \citenamefont {Bright}, \citenamefont {{van der Walt}}, \citenamefont
  {Brett}, \citenamefont {Wilson}, \citenamefont {Millman}, \citenamefont
  {Mayorov}, \citenamefont {Nelson}, \citenamefont {Jones}, \citenamefont
  {Kern}, \citenamefont {Larson}, \citenamefont {Carey}, \citenamefont {Polat},
  \citenamefont {Feng}, \citenamefont {Moore}, \citenamefont {{VanderPlas}},
  \citenamefont {Laxalde}, \citenamefont {Perktold}, \citenamefont {Cimrman},
  \citenamefont {Henriksen}, \citenamefont {Quintero}, \citenamefont {Harris},
  \citenamefont {Archibald}, \citenamefont {Ribeiro}, \citenamefont
  {Pedregosa}, \citenamefont {{van Mulbregt}},\ and\ \citenamefont {{SciPy 1.0
  Contributors}}}]{scipy}%
  \BibitemOpen
  \bibfield  {author} {\bibinfo {author} {\bibfnamefont {P.}~\bibnamefont
  {Virtanen}}, \bibinfo {author} {\bibfnamefont {R.}~\bibnamefont {Gommers}},
  \bibinfo {author} {\bibfnamefont {T.~E.}\ \bibnamefont {Oliphant}}, \bibinfo
  {author} {\bibfnamefont {M.}~\bibnamefont {Haberland}}, \bibinfo {author}
  {\bibfnamefont {T.}~\bibnamefont {Reddy}}, \bibinfo {author} {\bibfnamefont
  {D.}~\bibnamefont {Cournapeau}}, \bibinfo {author} {\bibfnamefont
  {E.}~\bibnamefont {Burovski}}, \bibinfo {author} {\bibfnamefont
  {P.}~\bibnamefont {Peterson}}, \bibinfo {author} {\bibfnamefont
  {W.}~\bibnamefont {Weckesser}}, \bibinfo {author} {\bibfnamefont
  {J.}~\bibnamefont {Bright}}, \bibinfo {author} {\bibfnamefont {S.~J.}\
  \bibnamefont {{van der Walt}}}, \bibinfo {author} {\bibfnamefont
  {M.}~\bibnamefont {Brett}}, \bibinfo {author} {\bibfnamefont
  {J.}~\bibnamefont {Wilson}}, \bibinfo {author} {\bibfnamefont {K.~J.}\
  \bibnamefont {Millman}}, \bibinfo {author} {\bibfnamefont {N.}~\bibnamefont
  {Mayorov}}, \bibinfo {author} {\bibfnamefont {A.~R.~J.}\ \bibnamefont
  {Nelson}}, \bibinfo {author} {\bibfnamefont {E.}~\bibnamefont {Jones}},
  \bibinfo {author} {\bibfnamefont {R.}~\bibnamefont {Kern}}, \bibinfo {author}
  {\bibfnamefont {E.}~\bibnamefont {Larson}}, \bibinfo {author} {\bibfnamefont
  {C.~J.}\ \bibnamefont {Carey}}, \bibinfo {author} {\bibfnamefont
  {{\.I}.}~\bibnamefont {Polat}}, \bibinfo {author} {\bibfnamefont
  {Y.}~\bibnamefont {Feng}}, \bibinfo {author} {\bibfnamefont {E.~W.}\
  \bibnamefont {Moore}}, \bibinfo {author} {\bibfnamefont {J.}~\bibnamefont
  {{VanderPlas}}}, \bibinfo {author} {\bibfnamefont {D.}~\bibnamefont
  {Laxalde}}, \bibinfo {author} {\bibfnamefont {J.}~\bibnamefont {Perktold}},
  \bibinfo {author} {\bibfnamefont {R.}~\bibnamefont {Cimrman}}, \bibinfo
  {author} {\bibfnamefont {I.}~\bibnamefont {Henriksen}}, \bibinfo {author}
  {\bibfnamefont {E.~A.}\ \bibnamefont {Quintero}}, \bibinfo {author}
  {\bibfnamefont {C.~R.}\ \bibnamefont {Harris}}, \bibinfo {author}
  {\bibfnamefont {A.~M.}\ \bibnamefont {Archibald}}, \bibinfo {author}
  {\bibfnamefont {A.~H.}\ \bibnamefont {Ribeiro}}, \bibinfo {author}
  {\bibfnamefont {F.}~\bibnamefont {Pedregosa}}, \bibinfo {author}
  {\bibfnamefont {P.}~\bibnamefont {{van Mulbregt}}},\ and\ \bibinfo {author}
  {\bibnamefont {{SciPy 1.0 Contributors}}},\ }\bibfield  {title} {\bibinfo
  {title} {{{SciPy} 1.0: Fundamental Algorithms for Scientific Computing in
  Python}},\ }\href {https://doi.org/10.1038/s41592-019-0686-2} {\bibfield
  {journal} {\bibinfo  {journal} {Nature Methods}\ }\textbf {\bibinfo {volume}
  {17}},\ \bibinfo {pages} {261} (\bibinfo {year} {2020})}\BibitemShut
  {NoStop}%
\bibitem [{\citenamefont {Jung}\ and\ \citenamefont
  {Rosch}(2007)}]{jungLowerBoundsConductivities2007}%
  \BibitemOpen
  \bibfield  {author} {\bibinfo {author} {\bibfnamefont {P.}~\bibnamefont
  {Jung}}\ and\ \bibinfo {author} {\bibfnamefont {A.}~\bibnamefont {Rosch}},\
  }\bibfield  {title} {\bibinfo {title} {Lower bounds for the conductivities of
  correlated quantum systems},\ }\href
  {https://doi.org/10.1103/PhysRevB.75.245104} {\bibfield  {journal} {\bibinfo
  {journal} {Physical Review B}\ }\textbf {\bibinfo {volume} {75}},\ \bibinfo
  {pages} {245104} (\bibinfo {year} {2007})}\BibitemShut {NoStop}%
\bibitem [{\citenamefont {Ye}\ \emph {et~al.}(2020)\citenamefont {Ye},
  \citenamefont {Machado}, \citenamefont {White}, \citenamefont {Mong},\ and\
  \citenamefont {Yao}}]{ye_emergent_2020}%
  \BibitemOpen
  \bibfield  {author} {\bibinfo {author} {\bibfnamefont {B.}~\bibnamefont
  {Ye}}, \bibinfo {author} {\bibfnamefont {F.}~\bibnamefont {Machado}},
  \bibinfo {author} {\bibfnamefont {C.~D.}\ \bibnamefont {White}}, \bibinfo
  {author} {\bibfnamefont {R.~S.~K.}\ \bibnamefont {Mong}},\ and\ \bibinfo
  {author} {\bibfnamefont {N.~Y.}\ \bibnamefont {Yao}},\ }\bibfield  {title}
  {\bibinfo {title} {Emergent hydrodynamics in nonequilibrium quantum
  systems},\ }\href {https://doi.org/10.1103/physrevlett.125.030601} {\bibfield
   {journal} {\bibinfo  {journal} {Physical Review Letters}\ }\textbf {\bibinfo
  {volume} {125}},\ \bibinfo {pages} {030601} (\bibinfo {year}
  {2020})}\BibitemShut {NoStop}%
\bibitem [{\citenamefont {Ye}\ \emph {et~al.}(2022)\citenamefont {Ye},
  \citenamefont {Machado}, \citenamefont {Kemp}, \citenamefont {Hutson},\ and\
  \citenamefont {Yao}}]{ye_universal_2022}%
  \BibitemOpen
  \bibfield  {author} {\bibinfo {author} {\bibfnamefont {B.}~\bibnamefont
  {Ye}}, \bibinfo {author} {\bibfnamefont {F.}~\bibnamefont {Machado}},
  \bibinfo {author} {\bibfnamefont {J.}~\bibnamefont {Kemp}}, \bibinfo {author}
  {\bibfnamefont {R.~B.}\ \bibnamefont {Hutson}},\ and\ \bibinfo {author}
  {\bibfnamefont {N.~Y.}\ \bibnamefont {Yao}},\ }\bibfield  {title} {\bibinfo
  {title} {Universal kardar-parisi-zhang dynamics in integrable quantum
  systems},\ }\href {https://doi.org/10.1103/PhysRevLett.129.230602} {\bibfield
   {journal} {\bibinfo  {journal} {Phys. Rev. Lett.}\ }\textbf {\bibinfo
  {volume} {129}},\ \bibinfo {pages} {230602} (\bibinfo {year}
  {2022})}\BibitemShut {NoStop}%
\bibitem [{\citenamefont {Barthel}(2013)}]{barthel2013}%
  \BibitemOpen
  \bibfield  {author} {\bibinfo {author} {\bibfnamefont {T.}~\bibnamefont
  {Barthel}},\ }\bibfield  {title} {\bibinfo {title} {Precise evaluation of
  thermal response functions by optimized density matrix renormalization group
  schemes},\ }\href {https://doi.org/10.1088/1367-2630/15/7/073010} {\bibfield
  {journal} {\bibinfo  {journal} {New Journal of Physics}\ }\textbf {\bibinfo
  {volume} {15}},\ \bibinfo {pages} {073010} (\bibinfo {year}
  {2013})}\BibitemShut {NoStop}%
\end{thebibliography}%

\appendix

\section{Comparison of correlators for $D(t)$ calculation}
\label{sec:correlator-comparison}
The time-dependent diffusion constant can be computed using multiple equivalent formulations relying on different time-dependent correlators: the energy-energy correlator, through Eq.~\ref{eq:Dt}, the current-current correlator, through Eq.~\ref{eq:current-grns}, and the energy-current correlator. Here we give a general derivation of this connection assuming spatial- and time-translation invariance. We use the charge $n$ instead of the energy for greater generality.

The time-dependent diffusion constant is defined as
\begin{align}
D(t) &= \frac 1 {2\nu} \frac \partial {\partial t} \int dx\, x^2 \langle n(x,t) n(0,0) \rangle
\end{align}
where $\nu=\langle n(x,t) n(x,t) \rangle$. Bringing the time derivative under the integral and using the conservation law gives
\begin{align}
D(t)&= \frac 1 {2\nu} \int dx\, x^2 \langle \dot n(x,t) n(0,0) \rangle \\
&= -\frac 1 {2\nu} \int dx\, x^2 \langle \nabla j(x,t) n(0,0) \rangle.
\end{align}
The resultant spatial derivative cancels out a factor of $x$ through integration by parts:
\begin{align}
D(t) &= -\frac 1 {2\nu} \int dx\, x^2 \nabla \langle j(x,t) n(0,0) \rangle \\
&= \frac 1 {\nu} \int dx\, x \langle j(x,t) n(0,0) \rangle.
\end{align}
This expression gives the time-dependent diffusion constant in terms of the charge-current correlator.

Applying this process again yields the current-current correlator. First, by spatial- and time-translation invariance, the charge-current correlator is equivalent to 
\begin{align}
D(t) = \frac 1 {\nu} \int dx\, x \langle j(0,0) n(-x,-t) \rangle.
\end{align}

which by the same procedure gives
\begin{align}
D(t) &= \frac 1 {\nu} \int_{0}^t dt' \, \frac d {dt'} \int dx\, x \langle j(0,0)  n(-x,-t') \rangle \\
&= \frac 1 {\nu} \int_{0}^t dt' \, \int dx\, \langle j(x,t) j(0,0) \rangle \\
&= \frac 1 {\nu L} \int_{0}^t dt' \, \langle J(x,t) J(0,0) \rangle.
\end{align}

\begin{figure}[t]
  \centering
  \includegraphics[width=\columnwidth]{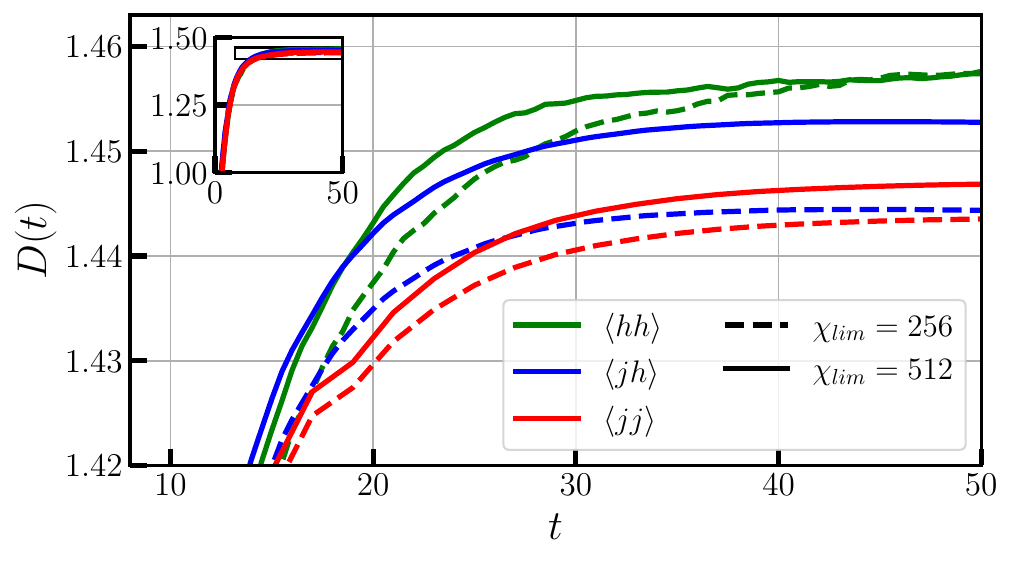}
  \caption{\label{fig:correlator_comparison} Comparison of time-dependent diffusion constant calculated using various correlators. Inset shows zoomed-out plot.}
\end{figure}

Using a DMT simulation, we calculate $D(t)$ using all three of these values and plot the generally consistent results in Fig.~\ref{fig:correlator_comparison}.
We consider $\expct{jj}$ in the rest of the paper in large part because it appears the best converged.

\section{Numerical Convergence in Trotter Step Size}\label{app:trotter} \label{s:trotter-convergence}
\begin{figure}[t]
  \centering
  \includegraphics[width=\columnwidth]{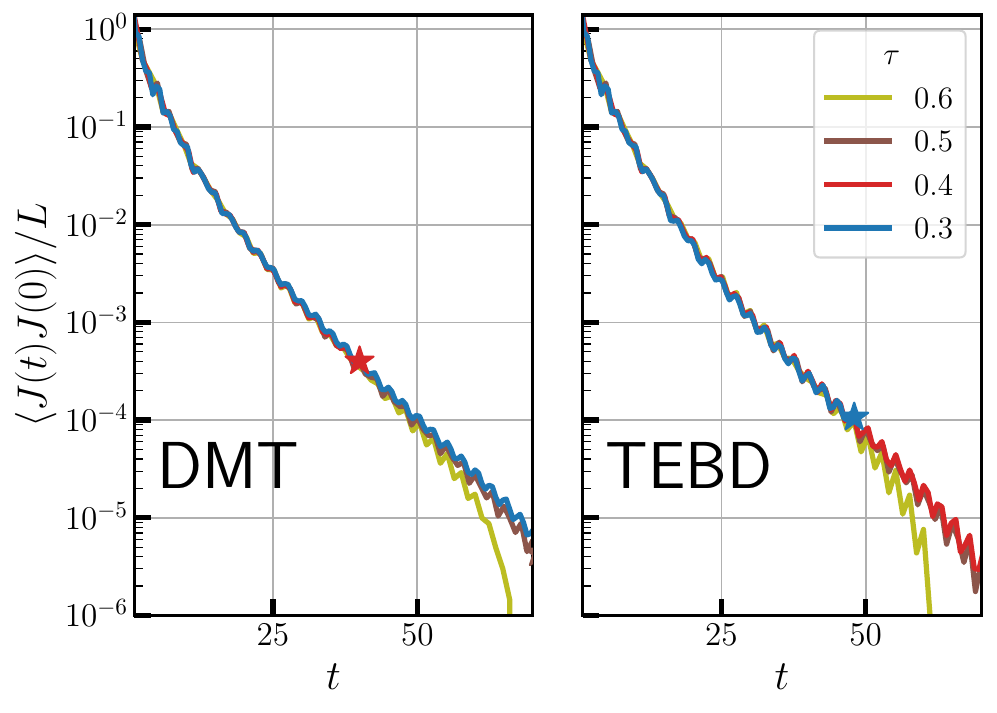}
  \caption{\label{fig:trotter-convergence} Current-current correlator decay for different Trotter step sizes $\tau$ using time-doubling. $\chi_\mathrm{lim}=512$. Stars mark the end of execution.}
\end{figure}

We recreate the current-current decay results (\cref{fig:dmt-tail-check}) for various Trotter step sizes $\tau$ to demonstrate convergence. The results are shown in \cref{fig:trotter-convergence}. Due to the specific Trotter algorithm used in this study (see \cref{sss:tebd}), we expect the error from each step to be $O(\tau^4)$ however, for long times these errors accumulate. We numerically find that all values up to $\tau = 0.5$ are convergent at times $t\gtrsim 40$.

\section{Time-Doubling} \label{s:time-doubling}
As described in \cref{ss:mpo-methods}, we use MPO methods to calculate the time-dependent operator $j_0(t) = e^{-iHt}j_0 e^{iHt}$ which relates to the current-current Green's function $\langle j_x(t) j_0(0) \rangle$ via \cref{eq:current-grns}. However under certain conditions---unitary evolution, infinite temperature, time translation invariance and spatial translation invariance---we can use a time-doubling trick~\cite{barthel2013} to effectively calculate $\langle j_x(2t)j_0(0) \rangle$ from the same MPO calculation.

To demonstrate this procedure, we rewrite the correlator as two time-evolved operators
\begin{equation}
  \langle j_x(2t) j_0(0) \rangle = \langle j_x(t) j_0(-t) \rangle.
\end{equation}
We calculate the first operator $j_x(t)$ by applying time-inversion and spatial translation onto the MPO $j_0(-t)$. Time inversion acts on the current operator as the complex conjugate, giving
\begin{equation}
j_x(t) = -\hat{T}(x)\, j^*_0(-t) \,\hat{T}(-x)\;.
\end{equation}
We use this approach in calculating the long-time portions of the MPO results.

\begin{figure}[t!]
  \centering
  \includegraphics[width=\columnwidth]{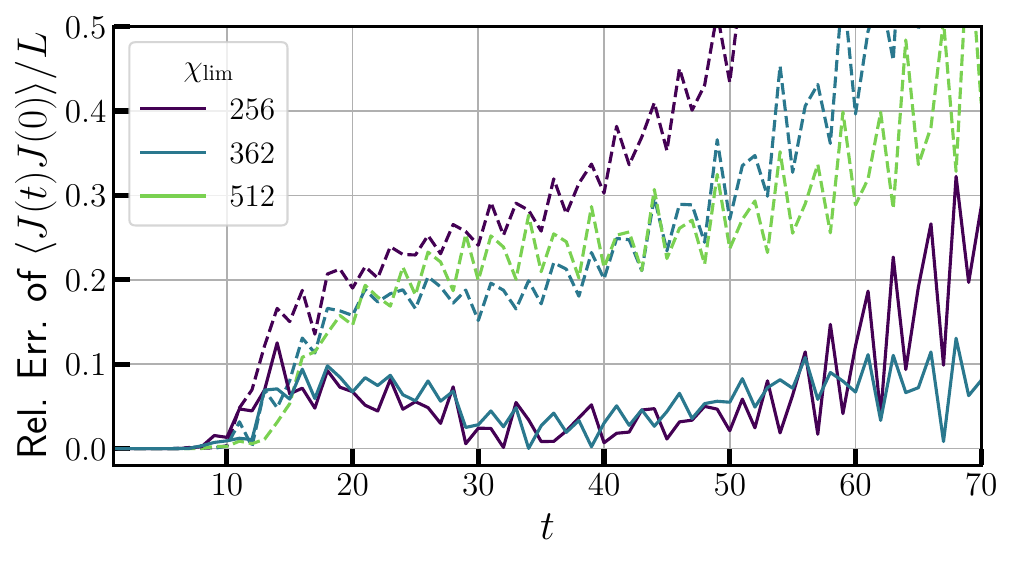}
  \caption{\label{fig:DMT-vs-TEBD-unsmoothed} Unsmoothed relative error in current-current correlator, computed by DMT (solid lines) and TEBD (dashed lines). See \cref{fig:DMT-vs-TEBD} for smoothed data and discussion.}
\end{figure}

\section{DMT for traceless operators}\label{app:dmt-details}
\subsection{As described in \onlinecite{white2018}}\label{app:dmt-prev}
DMT as described in \onlinecite{white2018} truncates a density matrix
without modifying any 3-site reduced density matrix.
Consider truncation of a matrix product operator $A(t)$ on a bond $j$ (separating sites $j,j+1$).
First, DMT as described in \onlinecite{white2018} uses a gauge transformation to form a correlation matrix
\begin{align*}\label{eq:Mdef}
  M_{\alpha\beta} = \tr \left[A(t) \hat y_{L\alpha} \hat y_{R\beta}\right], \quad 0 \le \alpha, \beta \le \chi - 1\;.
\end{align*}
$\chi$ is the bond dimension of the matrix product operator $A(t)$,
and $y_{L\alpha}, y_{R\beta}$ span the left and right Schmidt spaces at bond $j$, respectively.
The matrix $M$ serves as an orthogonality center:
the matrix product operator $A(t)$ can be written
\begin{align}
A(t) = \diag{4ex}{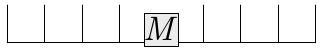}
\end{align}
where $\diag{2ex}{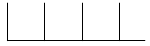}$ and $\diag{2ex}{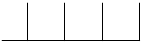}$ are left- and right-canonical, respectively.
The rank of this matrix $M$ therefore controls the bond dimension of the MPO,
and the MPO can be truncated by truncating the correlation matrix.

The gauge transformation giving the matrix $M_{\alpha\beta}$ is chosen so that the operators $\hat y_{L,\alpha}$ and $\hat y_{R,\beta}$ satisfy the following properties:
\begin{enumerate}
\item $\hat y_{L,\alpha}$, $\hat y_{R,\beta}$ are traceless for $\alpha,\beta> 0$
\item $\hat y_{L,\alpha}$, $\hat y_{R,\beta}$ are supported on sites $1, \dots, j$ and $j+1, \dots, L$
\item The image of $\hat \sigma^\mu_j$ in the bond space at bond $j$ is spanned by $\{\hat y_{L\alpha}\ :\ \alpha = 0, \dots, 3 \}$, i.e. 
  \begin{align}
    \tr \left[\hat \sigma^\mu_j \hat y_{L\alpha}\right] = 0,\quad 4 \le \alpha 
  \end{align}
  and likewise the image of $\hat \sigma^\mu_{j+1}$ in the bond space at bond $j$ is spanned by $\{\hat y_{R\beta}\ :\ \beta= 0, \dots, 3 \}$, i.e. 
  \begin{align}
    \tr \left[\hat \sigma^\mu_{j+1} \hat y_{R\beta}\right] = 0,\quad 4 \le \beta \;.
  \end{align}
\end{enumerate}
When the operator being truncated is a density matrix,
these properties of the $\hat y_{L,R}$ link the block structure of the matrix $M_{\alpha\beta}$ to local reduced density matrices:
one can check that modifications to $M_{\alpha,\beta}$ for $\alpha,\beta \ge 4$ do not change any three-site correlation functions.
When the operator being truncated is not a density matrix,
the quantities preserved are no longer reduced density matrices, but partial traces of the operator in question.

As described in \onlinecite{white2018}, DMT first forms the matrix of connected components 
\begin{subequations}
  \begin{align}\label{eq:corr-matrix-def}
  \tilde M_{\alpha\beta}
  &= \expct{\hat y_{L\alpha}\hat y_{R\beta}}
  - \expct{\hat y_{L\alpha}}\expct{\hat y_{R\beta}} \\
  &= M_{\alpha\beta} - \frac{ M_{\alpha 0} M_{0\beta}}{M_{00}}\label{eq:tildeM-construction}\;.
  \end{align}
\end{subequations}
It then truncates the submatrix $\tilde M_{\alpha\beta}|_{4 \le \alpha,\beta}$ by keeping leading principle components for a new matrix of connected components 
\begin{align}
  \label{eq:Mreduction}
  \tilde M' = 
  \left[
  \begin{array}{c|ccc}
    \quad&&&\\
    \hline\\
    &&XrP^{(\chi')}V&\\
    &&&\\
  \end{array}
  \right].
\end{align}
Finally it reassembles a new correlation matrix
\begin{align}
  M'_{\alpha\beta} &= \tilde M'_{\alpha\beta} + \frac{ M_{\alpha 0} M_{0\beta}}{M_{00}}\;.
\end{align}
from the truncated matrix $\tilde M'$ and the unchanged column and row $M_{\alpha 0},M_{0\beta}$.

\subsection{As modified for Heisenberg dynamics and used in this work}\label{app:dmt-mod}

Now consider Heisenberg evolution of a traceless operator $A$.
In constructing the matrix of connected components \eqref{eq:tildeM-construction},
the protocol of Sec.~\ref{app:dmt-prev} divides by the matrix element $M_{00}$.
When the operator being truncated is a density matrix, 
the matrix element $M_{00}$ encodes the trace of the density matrix.
But we are no longer evolving a density matrix: we are evolving a traceless operator $A$,
so dividing by $M_{00}$ is not natural.

Instead we truncate the matrix $M$ directly.
We use the same gauge transformation as in \cref{app:dmt-prev} and \onlinecite{white2018},
leading to operators $\hat y_{L\alpha},\hat y_{R\alpha}$ with the same properties.
The matrix
\begin{align}
  M_{\alpha\beta} = \tr \left[A \hat y_{L\alpha} \hat y_{R\beta} \right]
\end{align}
has a block structure analogous to that of the $M_{\alpha\beta}$ of \eqref{eq:Mdef};
in particular, modifications to $M_{\alpha\beta}$ for $\alpha,\beta \ge 4$ do not change the action of $A$ on any connected three-site region.

But now, instead of forming the matrix of connected components $\tilde M$ and truncating that,
we directly truncate the matrix $M$ for
\begin{align}
   M' = 
  \left[
  \begin{array}{c|ccc}
    \quad&&&\\
    \hline\\
    &&XrP^{(\chi')}V&\\
    &&&\\
  \end{array}
  \right].
\end{align}

\section{Additional convergence data} \label{s:additional-convergence}

\begin{figure}[htb]
  \centering
  \includegraphics[width=\columnwidth]{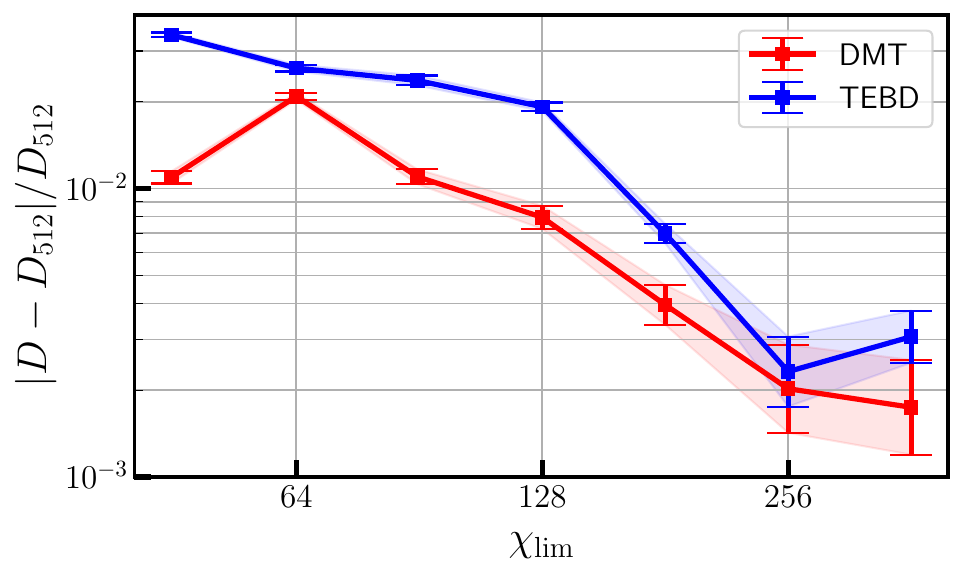}
  \caption{\label{fig:D_errors} Convergence of the diffusion coefficient with DMT and TEBD, calculated using $n=9$ Newton-Coates formula. Fine-grained integration ($\tau=0.1$) was performed up to $t=18$ with coarse grained ($\tau=0.5$) used afterwards. Errors are calculated by doubling the time step.}
\end{figure}

\begin{figure}[htb]
  \centering
  \includegraphics[width=\columnwidth]{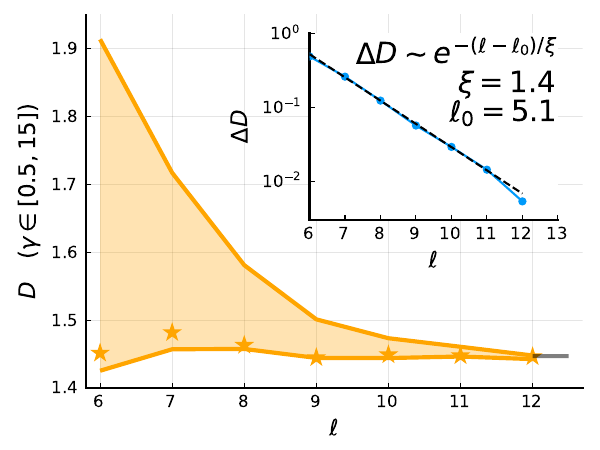}
  \caption{\label{fig:opgraph_deltaD} Maximum and minimum computed values of the diffusion coefficient in the OST dynamics for the range of $\gamma$ considered, $\gamma \in [0.5, 15]$. The maximum and minimum both converge to the DMT-computed value (denoted with a black mark) as $\ell$ is increased. For small $\ell$, the diffusion coefficient computed with $\gamma_{\text{semi-emp.}}$ (stars) is a good estimate. (Inset) The difference between the maximum and minimum $D$ over this interval of $\gamma$ shrinks exponentially in $\ell$, illustrating the exponentially increased insensitivity of $D(\gamma)$ to the noise rate $\gamma$.}
  \label{fig:opgraph_delta_D}
\end{figure}

We analyzed the convergence of the current-current correlator across time by comparing the results of the TEBD and DMT calculations with the largest bond dimension ($\chi_\mathrm{lim}=512$) DMT data.
In the main text, \cref{fig:DMT-vs-TEBD} contains data which is smoothed with a Savitzky-Gavoy filter. The corresponding unsmoothed data are shown here in \cref{fig:DMT-vs-TEBD-unsmoothed}. In each data series, the current exponentially decays with rates that are nearly the same and contains additional oscillatory modes that decay at a similar rate, as in \cref{eq:phenom-pole-structure}. The oscillatory part of the error is largely due to a mismatch in the amplitude and phase of the oscillations.

Additionally, we analyzed the convergence of the diffusion coefficient computed from the current-current correlator. \cref{fig:D_errors} shows the results for the diffusion coefficient computed with TEBD and DMT and various bond dimensions, subtracted from the result with the largest bond dimension DMT data. Roughly, the plot shows exponential improvement in the accuracy of the diffusion coefficient with increasing bond dimension; however, the pattern is not consistent enough to be conclusive.

Finally, we analyzed the convergence of the diffusion coefficient within the OST dynamics. \cref{fig:opgraph_deltaD} shows the range of predicted values for the diffusion coefficient when the absorption rate $\gamma$ is tuned across the entire range considered, $\gamma \in [0.5, 15]$. The size of this range shrinks exponentially with increasing $\ell_*$, and moreover converges onto the value computed using DMT calculations.

\section{Diffusion and the spectrum of the Liouvillian}\label{ss:diffusion-spectrum}

 In Sec.~\ref{ss:phenomenology} we related the diffusion coefficient to the time integral of the current-current correlator by a series of identities.
 We can understand how that relation fits into the broader long- but finite-wavelength behavior of the correlators $\expct{j_x(t)j_0(0)}$ and $\expct{\varepsilon_x(t) \varepsilon_0(0)}$
 by considering the pole structure of the Liouvillian of a large but finite system.\footnote{We take the system finite for convenience: so operators have discrete spectra. We also take the system to have periodic boundary conditions.}
Consider Fourier transforms of the energy and current operators
 \begin{align}
    \begin{split}
        \varepsilon_k &\propto \sum_x e^{-ikx} \varepsilon_x \\
        j_k &\propto \sum_x e^{-ikx} j_x\;.
    \end{split}
\end{align}
 The Liouvillian $\mathcal L$ is the linear operator generating a system's Heisenberg dynamics;
 it has a matrix element between general operators $A,B$
 \begin{align}
     \braket{A|\mathcal L | B} = i \tr\Big(A [H,B]\Big)\;,
 \end{align}
where $H$ is the system's Hamiltonian.
A correlation function $\expct{j_k(t)j_k(0)}$, then, is
\begin{align}
    C^{jj}(k,t) = \tr[j_k(t) j_k(0)] = \braket{j_k | e^{-i\mathcal L t}|j_k }\;.
\end{align}
Since the dynamics is unitary, the Liouvillian $\mathcal L$ is Hermitian. For any finite-sized system, the dynamics are thus described by a complete set of eigenvectors with real eigenvalues; this dynamics, however, includes recurrences in $C^{AB}(t)$ on timescales exponentially long in system size that do not occur in the thermodynamic limit.

For chaotic systems in the thermodynamic limit, correlations such as $\expct{j_k(t)j_k(0)}$ decay.
\cite{whiteEffectiveDissipationRate2021,von_keyserlingk_operator_2021,rakovszky2022} argue (with varying degrees of explicitness) that this should be understood in terms of a non-Hermitian effective Liouvillian $\mathcal L_\eff$ that matches the exact, Hermitian Liouvillian $\mathcal L$ for local operators.
DAOE~\cite{von_keyserlingk_operator_2021} and the OST dynamics of~\cite{whiteEffectiveDissipationRate2021} are two examples of such a non-Hermitian effective Liouvillian;
we give details of  the OST dynamics in Sec.~\ref{ss:liouv}.
Here we assume $\mathcal L_\eff$ is diagonalizable.
$\mathcal L_\eff$ again gives a correlation function $\braket{A|e^{-i\mathcal L_\eff t}| B}$, but now $\mathcal L_\eff$'s eigenvalues can have an imaginary part, leading to the observed decay.
In the long-time limit only the slowest eigenvalues---those with smallest imaginary part---remain.

The diffusive phenomenology of the previous section appears in these slowest eigenvalues
due to the constraints placed on the Liouvillian by translation invariance and the discrete continuity equation \eqref{eq:cl-continuity}.
Because the system is translation-invariant, the Liouvillian is block-diagonal in $k$ space;
call the blocks $\mathcal L_k$.
The continuity equation \eqref{eq:cl-continuity} implies that in the small-$k$ limit
\begin{align}
    \mathcal L_k \ket{\varepsilon_k} =  k \ket{j_k}\;.
\end{align}
In terms of orthonormal vectors $\ket{\hat j_k} =\|j_k\|^{-1} \ket{j_k}$ and $\ket{\hat \varepsilon_k} = \|\varepsilon_k\|^{-1} \ket{\varepsilon_k}$, where $\| \cdot\|$ is the Frobenius norm, the effective Liouvillian is
\begin{align} \label{eq:Lk-M}
    \mathcal L_k = a_k k \ketbra{\hat j_k}{\hat \varepsilon_k} + \mathrm{h.c.} + M_k
\end{align}
where $M_k$ acts on the space orthogonal to $\ket{\varepsilon_k}$, and $a_k = {\|j_k\|}/{\|\varepsilon_k\|}$ is a model-dependent constant.
{
In this notation the Laplace-space memory matrix $M_{\varepsilon\varepsilon}(z)$ of \onlinecite{jungLowerBoundsConductivities2007,lucasMemoryMatrixTheory2015} is
\begin{align}
\begin{split}
    M_{\varepsilon\varepsilon}(z) &= i\braket{j_k | (z + iM_k)^{-1} |j_k} \\
    &= ik^2\int_0^\infty dt\; e^{-zt} \braket{j_k| e^{-iM_kt} |j_k}\;.
\end{split}
\end{align}
That is, the memory matrix of \onlinecite{mori_transport_1965,zwanzig_memory_1961,jungLowerBoundsConductivities2007,lucasMemoryMatrixTheory2015}
is the Laplace transform of the correlation function generated by our $M_k$, which captures the dynamics of $j_k$ apart from the energy density.

For $k = 0$ the charge completely decouples from the rest of the dynamics, as expected from conservation; at small but nonzero $k$ the coupling can be treated perturbatively.
Expand $\mathcal L_{\eff,k}$ to leading order in $k$
\begin{align}
    \mathcal L_{\eff,k} = \mathcal L_{\eff,0} + k M'_0 + \Big(a_0 k \ketbra{\hat j_0}{\hat \varepsilon_0} + \mathrm{h.c.} \Big) + O(k^2)
\end{align}
and call the perturbation $V = M'_0 + (i a_0 \ketbra{\hat j_0}{\hat \varepsilon_0} + \mathrm{h.c.} ) $.
Using second-order perturbation theory, we calculate the eigenvalue corresponding to the charge $\varepsilon_k$:
\begin{align}\label{eq:pole-pert}
    \lambda_\varepsilon(k) = a_0^2 k^2 \braket{\hat j_0 | \mathcal L_{\eff,0}^{-1}| \hat j_0}.
\end{align}
This is the diffusive pole $z = D k^2$. We can calculate its coefficient by integrating the total current autocorrelation function:
\begin{align}
D &= \frac 1 {\nu L} \int_0^\infty dt\, \expct{J(t) J(0)}
\end{align}
noting that $a_0^2=\|J\|^2/\nu L$ from \eqref{eq:nu}.
This expression matches \eqref{eq:current-grns2} exactly.

\end{document}